# Neurite Exchange Imaging (NEXI): A minimal model of diffusion in gray matter with inter-compartment water exchange

Ileana O. Jelescu[1,2,3,*], Alexandre de Skowronski[1], Françoise Geffroy[4], Marco Palombo[5,6,7,#] and Dmitry S. Novikov[8,#]

[1]CIBM Center for Biomedical Imaging, Animal Imaging and Technology, Ecole Polytechnique Fédérale de Lausanne, Lausanne, Switzerland

[2]Department of Radiology, Lausanne University Hospital (CHUV), Lausanne, Switzerland

[3]School of Biology and Medicine, University of Lausanne (UNIL), Lausanne, Switzerland

[4]NeuroSpin, CEA, Saclay, France

[5]Cardiff University Brain Research Imaging Centre (CUBRIC), School of Psychology, Cardiff University, Cardiff, UK

[6]School of Computer Science and Informatics, Cardiff University, Cardiff, UK

[7]Centre for Medical Image Computing, Dept. of Computer Science, University College London, London, UK

[8]Center for Biomedical Imaging, Department of Radiology, New York University School of Medicine, New York, NY, USA.

*Corresponding author: Ileana Jelescu (ileana.jelescu@chuv.ch)

# These authors contributed equally to this work.

**Abstract**

Biophysical models of diffusion in white matter have been center-stage over the past two decades and are essentially based on what is now commonly referred to as the "Standard Model" (SM) of non-exchanging anisotropic compartments with Gaussian diffusion. In this work, we focus on diffusion MRI in gray matter, which requires rethinking basic microstructure modeling blocks. In particular, at least three contributions beyond the SM need to be considered for gray matter: water exchange across the cell membrane – between neurites and the extracellular space; non-Gaussian diffusion along neuronal and glial processes – resulting from structural disorder; and signal contribution from soma. For the first contribution, we propose Neurite Exchange Imaging (NEXI) as an extension of the SM of diffusion, which builds on the anisotropic Kärger model of two exchanging compartments. Using datasets acquired at multiple diffusion weightings ($b$) and diffusion times ($t$) in the rat brain *in vivo*, we investigate the suitability of NEXI to describe the diffusion signal in the gray matter, compared to the other two possible contributions. Our results for the investigated diffusion time window (~10-45 ms) show minimal diffusivity time-dependence and more pronounced kurtosis decay with time, which is well fit by the exchange model. Moreover, we observe lower signal for longer diffusion times at high $b$. In light of these observations, we identify exchange as the mechanism that best explains these signal signatures in both low-$b$ and high-$b$ regime, and thereby propose NEXI as the minimal model for gray matter microstructure mapping. We finally highlight multi-$b$ multi-$t$ acquisitions protocols as being best suited to estimate NEXI model parameters reliably. Using this approach, we estimate the inter-compartment water exchange time to be 15 – 60 ms in the rat cortex and hippocampus in vivo, which is of the same order or shorter than the diffusion time in typical diffusion MRI acquisitions. This suggests water exchange as an essential component for interpreting diffusion MRI measurements in gray matter.

1. **Introduction**

The bedrock of biophysical models of diffusion MRI is water compartmentalization. Morphologically, there are at least three compartments in brain tissue that are essential for interpreting an MRI measurement. The first one is a collection of micron-thin long and often branch-like cellular structures, referred to "cellular processes or projections" – either axons, dendrites or glial cell processes. There, water diffusion is locally unidirectional, and is typically modeled in terms of the so-called "sticks", i.e., zero-radius cylinders (Jespersen et al., 2007; Kroenke et al., 2004). The second compartment is cell bodies (soma), which are roughly spherical and of ~ 15μm in diameter (Palombo et al., 2021). Their size is generally comparable with a typical mean squared displacement (the diffusion length) of a water molecule during measurements. The third compartment is the extra-cellular space in which the first two are embedded.

In the white matter (WM), soma are typically neglected due to their relatively small density (5 – 10% ex vivo) (Andersson et al., 2020; Veraart et al., 2020). Furthermore, the myelin sheath around axons contributes to impermeability (i.e. negligible exchange with the extra-axonal water) over the diffusion MRI-relevant timescales, and thus the sought compartmentalization. Biophysical models of diffusion in WM have therefore gained a lot of traction and are essentially based on what is now commonly referred to as the "Standard Model" (SM) (Novikov et al., 2019, 2018a) of non-exchanging compartments with Gaussian diffusion: a collection of sticks (axons) with some orientation distribution function (ODF); an anisotropic extra-axonal space surrounding each local fascicle (a bundle of sticks) aligned in a given direction; and, if relevant, the "free water" compartment describing the partial volume contribution of the cerebrospinal fluid (CSF), free from the hindrances of the extra-axonal space. Within the SM family, a constellation of implementations has been proposed, each with its own acronym and its own further simplifying assumptions, e.g., on the shape of the ODF for fiber fascicles, or on the relations between the compartmental diffusivities and volume fractions (Fieremans et al., 2011; Jespersen et al., 2010; Novikov et al., 2018b; Reisert et al., 2017; Wang et al., 2011; Zhang et al., 2012). These models are widely used to characterize WM microstructure, and are occasionally applied in gray matter (GM). Physics beyond SM has been revealed in WM, such as the residual non-Gaussian diffusion along sticks (axons) (Arbabi et al., 2020; Fieremans et al., 2016; Lee et al., 2020a).

In this work, we focus on diffusion MRI in GM, which is sufficiently distinct from WM morphologically. This implies rethinking basic microstructure modeling blocks, leading to a different simplified picture of diffusion MRI-relevant microgeometry (Jelescu et al., 2020; Novikov, 2021; Palombo et al., 2020). In particular, at least three contributions beyond the SM need to be considered:

(i) exchange across the membrane of cellular processes;
(ii) non-Gaussian diffusion along cellular processes – resulting from structural disorder; and
(iii) signal contribution from cell bodies (soma).

For (i), as myelin content is limited in GM, there is growing evidence that water exchange across the neurite membrane cannot be neglected for typical clinical diffusion times (20 < *t* < 80 ms). Evidence for the deviation from the impermeable stick model for neurites in GM and its relationship to exchange has been highlighted in human cortex, with an estimated characteristic in vivo exchange time of 10 – 30 ms (Veraart et al., 2020, 2018a). Similar exchange time ranges have been reported for perfused neonatal mouse spinal cords (Williamson et al., 2019), while other groups have reported longer exchange times of 100 – 150 ms in astrocyte and neuron cultures (Yang et al., 2018), rat brain (Quirk et al., 2003) and rat brain cortical cultures (Bai et al., 2018). Much shorter exchange times (3 – 5 ms) have been recently reported in the rat brain ex vivo (J. L. Olesen et al., 2021a).

For (ii), the intra-compartment structural disorder along the effectively one-dimensional neurite has been suggested in the rat brain in vivo based on the power law exponent $\vartheta = ½$ (Novikov et al., 2014) in oscillating-gradient diffusion MRI data of (Does et al., 2003), at time scales (0.4 – 10 ms) long relative to the correlation length, but shorter than typical PGSE (Pulsed Gradient Spin Echo) diffusion times. Recently, a related kurtosis time-dependence with the same exponent, $K(t) \propto t^{-1/2}$, was found in the human cortex at long $t$ (Lee et al., 2020b). However, in that work, the inter-compartment exchange could not be ruled out (generating the sub-leading, faster kurtosis decay $K(t) \propto t^{-1}$ at long $t$), especially in the face of weak-to-absent time-dependent diffusivity within the diffusion time range accessible with PGSE on a clinical scanner (20 – 100 ms). Notably, structural disorder in the extra-cellular space is naturally provided by the embedded neurites, which, given the three-dimensional nature of diffusion in this compartment, would lead to diffusion and kurtosis time-dependence as $(\ln t)/t$ at long $t$ (Novikov et al., 2014).

Finally, for (iii), cell bodies (soma) occupy ~10 – 20% of gray matter by volume (Bondareff and Pysh, 1968; Motta et al., 2019; Shapson-Coe et al., 2021; Spocter et al., 2012) and may need to be modeled, as proposed in a recent three-compartment model (SANDI) that also accounts for this tissue component by representing soma as impermeable spheres (Palombo et al., 2020). The potential issue of inter-compartment exchange was partially circumvented by the use of short diffusion times ($t$ < 20 ms). Diffusion in soma was modeled in the Gaussian phase approximation, whereby diffusivity time-dependence in this compartment was retained, but higher-order terms (kurtosis and above) were neglected.

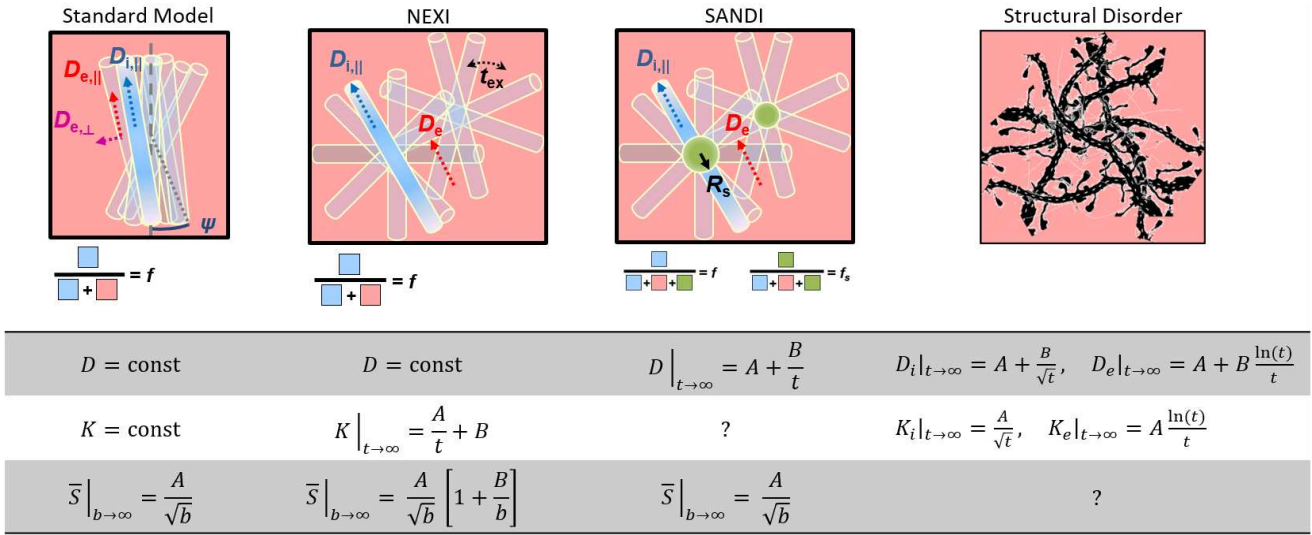

**Figure 1.** Sketch of relevant features and parameters in the Standard Model, NEXI, SANDI and structural disorder models, along with the associated functional form for time-dependence in diffusivity D and kurtosis K, as well as the functional form of the signal decay in the high-b regime. Note that parameters $A$ and $B$ stand for constants that are different in each instance. The **Standard Model** considers a collection of impermeable sticks – occupying a relative signal fraction $f$ – where diffusion is Gaussian and unidirectional with diffusivity $D_{i,\parallel}$ and an extra-neurite Gaussian anisotropic compartment with characteristic diffusivities $D_{e,\parallel}$ and $D_{e,\perp}$ parallel and perpendicular to the local neurite orientation, respectively. ODF anisotropy can be characterized by the $l=2$ order rotational invariant $p_2$ or its derived dispersion angle $c_2 \equiv \langle cos^2\psi \rangle = \frac{2p_2+1}{3}$. **NEXI** considers a collection of randomly-oriented sticks – occupying a relative signal fraction $f$ – where diffusion is unidirectional with diffusivity $D_{i,\parallel}$ and an extra-neurite Gaussian isotropic compartment with characteristic diffusivity $D_e$. The two compartments exchange with a characteristic time $t_{ex}$. **SANDI** considers a similar picture as NEXI, but accounts for a third compartment of spheres of radius $R_s$ (occupying a relative fraction $f_s$) and neglects inter-compartment exchange. The **structural disorder** model assumes a certain type of disorder (here short-range disorder) and its signature in diffusion and kurtosis time-dependence, also as a function of the spatial dimensionality in which the disorder is manifest (1D for intra-neurite water, 2D or 3D for extracellular water).

**Figure 1** summarizes the picture underlying each model, the expected functional forms of time-dependent diffusion and kurtosis in the long-time limit as well as the functional form of the powder-averaged signal at high $b$. Model details and full expressions of the functional forms are provided in the Theory Section.

This work is organized in two main parts.

First, we examine the diffusion and kurtosis time-dependence as well as powder-average signal signature at high *b*-value in the rat brain *in vivo* in order to assess the importance of the effects (i) – (iii) above. Within the ranges of diffusion weightings (*b*) and diffusion times (*t*) explored, we highlight negligible diffusivity time-dependence and a time-dependent kurtosis which can be consistent with the $t^{-1}$ power-law, both trends compatible with exchange, whereby structural disorder may be less relevant in modeling rat cortex in this diffusion time range. Signal decay curves at high *b*-values and for different diffusion times are also better accounted for by an exchange term as compared with adding the soma compartment, as also recently shown in ex vivo rat cortex (J. L. Olesen et al., 2021a).

Second, we therefore propose NEXI (Neurite EXchange Imaging), an implementation of the anisotropic Kärger model of exchange as the minimal model suitable for gray matter, accounting only for the effect (i) above. We show in simulations that the availability of multi-*b* multi-*t* data is critical for the reliable estimation of model parameters, in particular of the exchange time. We further demonstrate NEXI performance in experimental data. Fit stability permitting, the model could be extended to account for a soma fraction as well (J. L. Olesen et al., 2021a; Palombo et al., 2020).

For our experiments, we exploit the potential of strong preclinical gradients (1 T/m) to probe a range of short to intermediate diffusion times (10 – 45 ms) with strong diffusion-weighting (up to *b* = 10 ms/μm²). The GM regions of interest (ROIs) are cortex and hippocampus, with corpus callosum, internal capsule and cingulum serving as reference WM ROIs. We take advantage of the rat brain anatomy where cortex is much less affected by partial volume effects with neighboring white matter or CSF than its human counterpart.

## 2. Methods
### 2.1. Theory

Let us first provide a theoretical description of the models sketched in **Figure 1**, that will be compared throughout this work.

**2.1.1. Standard Model**. To build a GM model, we begin with the Standard Model (Novikov et al., 2019) of brain tissue composed of two compartments (we will neglect the CSF contribution in what follows, assuming voxels are devoid of the CSF contamination). In SM, the intra-neurite compartment – occupying a relative signal fraction $f$ – is modeled as a collection of "sticks" (zero-radius cylinders) where diffusion is unidirectional with diffusivity $D_{i,\parallel}$. The extra-neurite compartment – the immediate environment of sticks – is modeled as a Gaussian anisotropic medium with characteristic diffusivities $D_{e,\parallel}$ and $D_{e,\perp}$ parallel and perpendicular to the local orientation of a neurite fascicle, respectively. For the fascicle oriented along the unit direction **n**, its signal (response) in the unit direction **g** is:

$$\mathcal{K}_{SM}(q, t, \mathbf{g} \cdot \mathbf{n}; f, D_{i,\parallel}, D_{e,\parallel}, D_{e,\perp}) = f e^{-q^2 t D_{i,\parallel}(\mathbf{g}\cdot\mathbf{n})^2} + (1-f) e^{-q^2 t \left(D_{e,\parallel}(\mathbf{g}\cdot\mathbf{n})^2 + D_{e,\perp}(1-(\mathbf{g}\cdot\mathbf{n})^2)\right)} \quad (1)$$

The signal attenuation (1) for the elementary fascicle is then convolved on the unit sphere $|\mathbf{n}| = 1$ with the orientation distribution function (ODF) $P(\mathbf{n})$ for the neurites to give the overall SM signal

$$S_{SM}(q, \mathbf{g}, t) = \int \mathcal{K}_{SM}(q, t, \mathbf{g} \cdot \mathbf{n}; f, D_{i,\parallel}, D_{e,\parallel}, D_{e,\perp}) P(\mathbf{n}) \, \mathrm{d}\mathbf{n}. \quad (2)$$

The SM is suitable for white matter, where neurites – mainly myelinated axons – are the dominant structure (vs negligible soma), and the accepted assumption is to neglect the exchange between intra- and extra-neurite spaces due to the myelin sheath.

**2.1.2. NEXI: adding exchange to SM.** In the GM, most neurites are unmyelinated and inter-compartment water exchange across the cell membrane may be non-negligible for clinical diffusion times ($t > 20$ ms). A model of two exchanging compartments, one being the collection of isotropically oriented neurites (sticks) and the other being the extra-neurite space, is built based on the anisotropic Kärger model for a coherent fiber tract (Fieremans et al., 2010; Kärger, 1985). Namely, the signal (response) from an elementary coherent ensemble of "neurite + its proximal extracellular space" is the result of mixing two anisotropic Gaussian compartments in Eq (1) by the barrier-limited exchange, with the rates $r_{ie}$ and $r_{ei}$ related by the detailed balance condition $fr_{ie} = (1-f)r_{ei}$:

$$\mathcal{K}(q,t,\mathbf{g}\cdot\mathbf{n}; f, D_{i,\|}, D_{e,\|}, D_{e,\perp}, t_{ex}) = f'e^{-q^2 t D_i'} + (1-f')e^{-q^2 t D_e'} \tag{3}$$

$$D_{i/e}' = \frac{1}{2}\left\{D_i + D_e + \frac{1}{q^2 t_{ex}} \mp \left[\left[D_e - D_i + \frac{2f-1}{q^2 t_{ex}}\right]^2 + \frac{4f(1-f)}{q^4 t_{ex}^2}\right]^{\frac{1}{2}}\right\} \tag{4}$$

$$f' = \frac{1}{D_i' - D_e'}[fD_i + (1-f)D_e - D_e'] \tag{5}$$

where $D_i \equiv D_{i,\|}(\mathbf{g}\cdot\mathbf{n})^2$, $D_e \equiv D_{e,\|}(\mathbf{g}\cdot\mathbf{n})^2 + D_{e,\perp}(1-(\mathbf{g}\cdot\mathbf{n})^2)$ and $t_{ex} = 1/r$, with $r = r_{ie} + r_{ei}$. In other words, the bi-exponential expression in Eq (3) is reminiscent of the SM expression in Eq (1), but with *apparent* fractions and diffusivities that depend on all model parameters as well as on diffusion wavenumber $q$ and on the characteristic exchange time $t_{ex}$.

In principle, the response function (3) should be convolved with the GM neurite ODF to get the overall signal in direction **g**, as in Eq (2). Since diffusion anisotropy in GM can be variable across cortical layers, though overall negligible (FA~0.15), we choose to consider the orientational average (the so-called powder average) of the signal instead, which is independent of the ODF. By the same token, we further assume the extra-neurite space to be isotropic $D_{e,\|} = D_{e,\perp} \equiv D_e$. This approximation helps reduce the number of parameters to be estimated and raises the precision on the remaining ones. The model parameters are therefore $\mathbf{p} = [f, D_{i,\|}, D_e, t_{ex}]$ (**Figure 1**). Hence, the model signal equation, to be fit to the powder-averaged measured signal $\bar{S}$, is:

$$\bar{S}(q,t) = S|_{q=0} \cdot \int_0^1 \mathcal{K}(q,t,\mathbf{g}\cdot\mathbf{n}; \mathbf{p})d(\mathbf{g}\cdot\mathbf{n}) \tag{6}$$

Technically, this means that we are only using the $l=0$ rotational invariant of the overall directional signal, and discarding potential information residing in $l=2,4,\ldots$ invariants. Nonetheless, the latter are small in GM, and are not expected to contribute much information to the original directional signal.

We call this implementation, Eqs (3)–(6), NEXI.

**NEXI kurtosis.** One of the assumptions behind the Kärger model, and thus behind NEXI, is time-independent diffusivity. If this condition is met, the NEXI kurtosis of Eq (6) (see Appendix for the derivation) is:

$$K(t) = K_0 \cdot \frac{2t_{ex}}{t}\left[1 - \frac{t_{ex}}{t}\left(1 - e^{-t/t_{ex}}\right)\right] + K_\infty. \tag{7}$$

It has two contributions: the inter-compartment heterogeneity $\sim K_0$ which decays to zero as $1/t$ at long times $t \gg t_{ex}$ as a result of exchange, identical to that found in (Fieremans et al., 2010; Jensen et al., 2005); and the constant offset $K_\infty$ that captures potential residual sources of kurtosis in the limit $t \gg t_{ex}$. Residual kurtosis could stem from partial volume (macroscopic heterogeneity) within a voxel (in our experiment, this is deemed small due to homogeneity of rat GM at our imaging resolution) and, in the case of powder-averaged signal, from non-exchanging microscopic anisotropic structures (Szczepankiewicz et al., 2016). Indeed, the powder-averaged signal, Eq (6), has residual kurtosis, as derived in the Appendix; it originates from the assumption that exchange happens only within a local neighborhood, between a single stick (neurite) and its accompanying extracellular space. The actual tissue geometry may differ from this latter assumption. If multiple sub-units of "neurite + extracellular space" with different orientations fall within the same volume $\sim L^3(t)$, where $L \sim \sqrt{6D_e t}$ is the diffusion length, then exchange can occur across multiple neurites. Since the extra-cellular space is a connected medium, at long times $t \gg t_{ex}$, exchange would coarse-grain the medium fully, yielding $K_\infty = 0$.

We note that the absence of exchange across multiple neurites is a mathematical construct to make the problem solvable from an analytical point of view. A plausible picture is naturally one with exchange across multiple neurites, given the random orientations of neurites and the tight network they constitute. This plausible picture led us to assume isotropic ECS diffusion and $K_\infty = 0$. At reasonably short diffusion times as the ones used here, the two pictures are compatible, but for longer diffusion times, exchange across multiple neurites is of course expected.

Finally, we also note that even for anisotropically oriented neurites, the kurtosis of the non-powder-averaged signal will have the functional form of Eq (7), with the parameters $K_0$ and $K_\infty$ depending on the ODF but the exchange time being ODF-independent. Hence, we will use Eq (7) to analyze mean kurtosis. Fitting Eq (7) to the measured mean kurtosis can provide a complementary estimate of $t_{ex}$ stemming from the low-$b$ regime (within the convergence radius of the cumulant expansion), to be compared to the one from the full NEXI model (6). We label this estimate $t_{ex}^{K(t)}$.

**High-$b$ scaling.** An independent hallmark of the model (3)–(6) is the functional form

$$\overline{S}|_{b\to\infty} = \sqrt{\frac{\pi}{4}} \frac{f}{\sqrt{bD_{i,\parallel}}} e^{-\frac{t(1-f)}{t_{ex}}} \left[1 + \frac{2(1-f)\,t/t_{ex} + f(1-f)(t/t_{ex})^2}{bD_{e,\perp}} + O\left(\frac{1}{b^2}\right)\right] \tag{8}$$

of its expansion in the inverse powers of the diffusion weighting parameter $b = q^2 t$. The first term in the square brackets, corresponding to $b^{-1/2}$ decrease, is the signature of impermeable sticks (Callaghan et al., 1979; McKinnon et al., 2017; Veraart et al., 2019). The subsequent $b^{-3/2}$ term arises due to slow exchange, such that $t/t_{ex} \ll bD$, where $D$ is the smallest of the compartment diffusivities; the lower bound is practically set by $D_{e,\perp} = D_e$. The $\sim t/t_{ex}$ term was obtained by (Veraart et al., 2020), and the $\sim (t/t_{ex})^2$ term by (J. L. Olesen et al., 2021a).

We note that the Kärger Model is treated in the narrow pulse approximation regime. This condition translates into $\delta \ll t_{ex}$, which will be justified a posteriori by comparing experimental values of $\delta = 4 - 4.5$ ms and $t_{ex} = 20 - 40$ ms. However, a numerical solution to the Kärger model in the finite pulse regime (J. L. Olesen et al., 2021a) could be implemented in the case of longer diffusion pulses (or shorter exchange times).

**2.1.3. Structural disorder**. The assumption of Gaussian compartments may break in the presence of irregularities on length scales similar to the diffusion length, such as dendritic spines and neurite beading. While there is no analytical formula to describe the signal in this case exactly, the relative importance of non-Gaussian effects can be determined by examining the diffusivity and kurtosis time-dependence at diffusion times $t \gg t_c$, the time $t_c$ to diffuse past the disorder correlation length (Novikov et al., 2014). In particular, kurtosis should follow a $t^{-1/2}$ functional form in the case of 1-dimensional disorder (Dhital et al., 2018; Lee et al., 2020b):

$$K(t)|_{t \gg t_c} \simeq A \cdot t^{-\vartheta} + K_\infty, \qquad \vartheta = 1/2 \tag{9}$$

or $(\ln t)/t$ in the case of 2-dimensional disorder (Burcaw et al., 2015; Lee et al., 2020b). The offset $K_\infty$ arises in the case of residual voxel heterogeneity in the long-time limit, similar to that in Eq (7). If the relevant correlation time $t_c$ for diffusion across these structural irregularities is of the order of $t_{ex}$, the competing effects of coarse-graining over the structural disorder and of exchange are both contributing significantly to the time-dependence of the measured $K(t)$, which complicates the interpretation (Lee et al., 2020b).

**2.1.4. SANDI: adding soma.** A three-compartment model (SANDI) was proposed as an extension of the SM that models the total direction-averaged signal as the sum of three non-exchanging compartments (Palombo et al., 2020): (i) randomly oriented sticks with intra-stick axial diffusivity $D_{i,\parallel}$ and relative signal fraction $f$; (ii) restriction in sphere of apparent radius $R_s$, fixed intra-sphere diffusivity $D_s = 3\frac{\mu m^2}{ms}$ and relative signal fraction $f_s$ (modeled in the Gaussian phase approximation); (iii) Gaussian isotropic diffusion in the extracellular space with diffusivity $D_e$ and relative signal fraction $f_e = 1 - f - f_s$. SANDI provides estimates for the five model parameters: [$f$, $f_s$, $D_{i,\parallel}$, $D_e$, $R_s$] which by design should be independent of diffusion time. The direction-averaged SANDI signal is:

$$\frac{\bar{S}(b)}{S(0)} = f \cdot \sqrt{\frac{\pi}{4}} \frac{1}{\sqrt{bD_{i,\parallel}}} \text{erf}\left(\sqrt{bD_{i,\parallel}}\right) + f_s \cdot \bar{A}_s(b, D_s, R_s) + f_e \cdot e^{-bD_e} \qquad (10)$$

where

$$\bar{A}_s(b, D_s, R_s) \approx \exp\left\{-\frac{2g^2 R_s^4}{D_s} \sum_{m=1}^{\infty} \frac{\alpha_m^{-4}}{\alpha_m^2 - 2} \right.$$
$$\left. \cdot \left[2\delta - \frac{R_s^2}{\alpha_m^2 D_s}\left(2 + e^{-\alpha_m^2 D_s (\Delta-\delta)/R_s^2} - 2e^{-\alpha_m^2 D_s \delta/R_s^2} - 2e^{-\alpha_m^2 D_s \Delta/R_s^2} + e^{-\alpha_m^2 D_s (\Delta+\delta)/R_s^2}\right)\right]\right\}$$

with δ and Δ the diffusion gradient pulse duration and separation, respectively, g the product of diffusion gradient amplitude and gyromagnetic ratio, $\alpha_m$ the m-th root of $\frac{1}{2}J_{\frac{3}{2}}(\alpha) = \alpha J'_{\frac{3}{2}}(\alpha)$, and $J_n(x)$ the Bessel function of the first kind. In practice, summation up to m=20 roots is sufficient for a good approximation.

### 2.2. Experimental

Animal experiments were approved by the Service for Veterinary Affairs of the canton of Vaud. Six Wistar rats (Charles River) weighing 250 - 300g were scanned on a 14T Bruker system equipped with 1 T/m gradients (Resonance Research Inc.) using a home-built surface quadrature transceiver. Rats were set up and maintained under isoflurane anesthesia, and body temperature was monitored and maintained around 38°C for the duration of the experiment. Diffusion MRI data were acquired using a PGSE EPI sequence, with parameters provided in **Table 1**. All six datasets were used to assess the behavior of time-dependent diffusion and kurtosis, while four datasets (labeled 1 – 4) were used for high-*b* signal analysis, SANDI and NEXI estimations. In datasets 5 – 6 we prioritized a larger number of diffusion times over *b*-values to capture trends in *D*(*t*) and *K*(*t*).

Images were denoised using MP-PCA and corrected for Rician bias (Veraart et al., 2016b, 2016a), for Gibbs ringing (Kellner et al., 2016) and for motion (Jenkinson et al., 2002). No strong distortions due to eddy currents were observed.

Regions of interest (ROI) in both white matter – internal capsule (IC), corpus callosum (CC) and cingulum (Cg) – and gray matter – cortex (CTX) and hippocampus (HPC) – were manually drawn.

| # Datasets | 3 | 1 | 1 | 1 |
|---|---|---|---|---|
| Dataset Label | 1 – 3 | 4 | 5 | 6 |
| TE (ms) | 50 | 58 | 52 | 58 |
| δ (ms) | 4.5 | 4.5 | 4 | 4 |
| Δ (ms) | 12, 20, 30, 40 | 11, 25, 45 | 10, 15, 20, 25, 30, 40 | 10, 15, 20, 25, 30, 35, 40, 45 |
| $b$-values (ms/um$^2$) | 1, 2.5, 4, 5.5, 7, 8.5, 10 | 1, 2.5, 5, 6, 7, 8, 9, 10 | 1, 1.8, 2.5 | 1, 1.4, 2.5 |
| Dirs. per shell | 24 | 24 | 24 | 24 |
| TR (ms) | 2500 | 3000 | 2500 | 3000 |
| In-plane res (mm$^2$) | 0.2 x 0.2 | 0.25 x 0.25 | 0.2 x 0.2 | 0.25 x 0.25 |
| Slice thickness (mm) | 0.5 | 0.8 | 0.5 | 0.8 |

**Table 1.** Acquisition parameters for the six datasets included in this study.

### 2.3. Impact of each contribution (i)-(iii): exchange, structural disorder & soma

**Standard Model**: To examine potential time-dependence and $b$-range dependence of SM estimates in various brain regions, for the four datasets with $b_{max}$ = 10 ms/µm$^2$, the SM parameters were estimated for each diffusion time using likelihood maximization in the rotational invariant framework RotInv using up to $l = 4$ (Novikov et al., 2018b) on various data subsets ($b_{max}$ =2.5, 6 or 10). RotInv was implemented in Matlab using non-linear least-squares minimization with a trust-region-reflective algorithm ('lsqnonlin' function). The $D_{i,\parallel} > D_{e,\parallel}$ solution was favored by choosing a random algorithm initialization that met this inequality. The time- and $b_{max}$- dependence of model parameters were evaluated in the various brain ROIs. Notable time-dependence of SM parameter estimates was tested via the slope – and its uncertainty – of a simple linear regression.

**Time-dependent diffusion and kurtosis**: To determine the extent to which inter-compartment exchange and/or structural disorder are relevant in various brain regions, diffusion and kurtosis tensors were estimated for each diffusion time using shells up to $b$ = 2.5 ms/µm$^2$ and a weighted linear least-squares algorithm custom-written in Matlab (Veraart et al., 2013), from which mean diffusivity and kurtosis were derived. The time-dependence of these metrics was evaluated in the various brain ROIs. Notable time-dependence of diffusivities was first tested via the slope – and its uncertainty – of a simple linear regression. To establish the dominant power-law of $K(t)$ decay, Eq (9) with variable $\vartheta$ was fit to the experimental $K(t)$ using the 'lsqnonlin' function a trust-region reflective algorithm in Matlab. The Kärger time-dependent kurtosis, Eq (7), was also fit to the experimental $K(t)$, either allowing for nonzero $K_\infty$ or setting it to zero. The 1D structural disorder functional form (Eq (9) with $\vartheta = 1/2$) was also fit to the measured $K(t)$ for comparison.

For SM estimates, as well as for $D(t)$ and $K(t)$, the uncertainty on fit parameters was estimated using a bootstrapping method where random noise with variance equal to the residual variance was added to the datapoints for N=1000 realizations, from which mean and standard deviation of the estimated parameters were extracted.

**Soma vs exchange**: The signal was averaged over each shell $\bar{S}(b,t)$ to fit either the SANDI model, Eq (10) (Palombo et al., 2020), for each diffusion time *t* separately, or the NEXI model, for all diffusion times *t* jointly (see Section 2.4 on NEXI parameter estimation below).

The SANDI fit was performed using its implementation in the accelerated microstructure imaging via convex optimization (AMICO) framework in Python 3.5 (Daducci et al., 2015), publicly available at: https://github.com/daducci/AMICO/wiki/Fitting-the-SANDI-model. Briefly, AMICO (Daducci et al., 2015) rewrites Eq (10) as a linear system *A***x**=**y**, where *A* = [$A_{stick}$, $A_{sphere}$, $A_{extra}$] is a matrix whose columns contain simulated signals of each compartment (stick, sphere or isotropic Gaussian), **y** is the vector of measured signals, and **x** the unknown contributions. To build *A*, we used a dictionary of signals simulated using: 5 values of $D_{i,\parallel}$ linearly spaced within the interval [0.25, 3] μm²/ms (namely **p**$_{stick}$) for $A_{stick}$; 5 values of $R_s$ linearly spaced within the interval [1, 12] μm (namely **p**$_{sphere}$) for $A_{sphere}$; and 5 values of $D_e$ linearly spaced within the interval [0.25, 3] μm²/ms (namely **p**$_{extra}$) for $A_{extra}$. The elements of **x** are then estimated using non-negative least squares with Tikhonov regularization (Efron et al., 2004) (regularization parameter $\lambda_2$ = 0.005) using the Lasso function implemented in the SPAMS optimization toolbox (http://spams-devel.gforge.inria.fr). From **x**, we then computed the SANDI model parameters as: $f = \frac{\sum_{i=1}^{5} x_i}{\sum_{i=1}^{15} x_i}$; $f_s = \frac{\sum_{i=6}^{10} x_i}{\sum_{i=1}^{15} x_i}$; $D_{i,\parallel} = \frac{\sum_{i=1}^{5} x_i p_{stick,i}}{\sum_{i=1}^{5} x_i}$; $R_s = \frac{\sum_{i=6}^{10} x_i p_{sphere,i}}{\sum_{i=6}^{10} x_i}$; $D_e = \frac{\sum_{i=11}^{15} x_i p_{extra,i}}{\sum_{i=11}^{15} x_i}$. The fit provided estimates for the five model parameters: [$f$, $f_s$, $D_{i,\parallel}$, $R_s$, $D_e$] in the cortex and hippocampus at each investigated diffusion time. The dependence of SANDI model parameters on diffusion time was quantified by computing the mean percentage difference of parameter estimates at each time with respect to the shortest diffusion time (*t* = 12 ms), and, in parallel, by performing one-way ANOVA as a function of time and reporting the significant differences pairwise for available diffusion times. Finally, the significance of a linear trend of model parameters over time was also calculated.

The SANDI parameter estimates at the shortest diffusion time were also used to predict the signal decay in the cortex at longer diffusion times, and compared to experimental outcomes, as suggested in (J. L. Olesen et al., 2021b).

The NEXI fit was performed using a non-linear least-squares (NLLS) optimization based on a quasi-Newton algorithm without constraints, implemented as 'fminunc' function in Matlab.

The performance of SANDI and NEXI to capture the deviation from the stick model at high *b*-values and the qualitative signal decay curves across multiple diffusion times was evaluated and compared.

### 2.4. NEXI parameter estimation

**Simulations:** Synthetic signals (*N* = 10⁴) were generated based on Eqs (3) – (6) assuming a protocol of *b* = 0 and seven shells at *b* = 1, 2.5, 4, 5.5, 7, 8.5 and 10 ms/μm², four diffusion times (*t* = 12, 20, 30, 40 ms) and a realistic SNR level of 100 – as estimated from experimental data in cortex following MP-PCA denoising and powder-averaging – see *Experimental* paragraph below. The ground truth was either fixed to $[t_{ex}, D_{i,\parallel}, D_e, f] = [20, 2.5, 0.75, 0.34]$ with only the noise realization changing for each iteration, or randomly chosen within physical ranges, that is $t_{ex} \in [5, 120]$, $D_{i,\parallel} \in [1.5, 3]$, $D_e \in [0, 1.5]$ and $f \in [0.1, 0.9]$, thus enforcing the $D_{i,\parallel} > D_e$ solution of the NEXI model. The exploration of disjoint intervals with $D_{i,\parallel} > D_e$ was supported by experimental data where, when repeating the NEXI estimation by varying the algorithm initialization (within full ranges $D_{i,\parallel}, D_e \in [0, 3]$) the mode of the outcome distribution yielded $D_{i,\parallel} > D_e$ in both cortex and hippocampus (see Results and **Figure S8**).

**Parameter estimation** was done either based on the signals for each diffusion time separately (as in standard multi-shell datasets) or jointly. A widespread non-linear least-squares (NLLS) minimization algorithm was used to estimate model parameters. NLLS used a trust-region-reflective algorithm with box constraints for multi-shell data and quasi-Newton algorithm without constraints for multi-shell multi-$t$ data. We also tested the performance of a deep learning (DL) algorithm for parameter estimation, in terms of precision and accuracy with respect to the more widespread NLLS approach, in the perspective of providing a fast implementation of NEXI with on-the-fly estimation of model parameter maps (Supplementary Methods).

**Impact of *b* range**. In the case of the joint diffusion times fit, we further evaluated the impact of the maximum *b*-value on NEXI estimates by retaining only subsets of the data for the estimation: $b_{max}$ = 2.5 (2 shells), 5.5 (4 shells), or 10 ms/µm² (7 shells). For completeness, we also compared the performance of the estimation for datasets with variable $b_{max}$ but same number of equally spaced shells ($N_{shells}$ = 7).

**Experimental**: The signal was averaged over each shell $\bar{S}(b,t)$ to be used for NEXI parameter estimation. All four model parameters were estimated with NLLS (and DL) by using all shells and diffusion times jointly. This joint fit was performed on a voxel-wise basis to generate parametric maps. The $t_{ex}$ estimate from NEXI was also compared to $t_{ex}^{K(t)}$.

The NLLS fit was unconstrained. To test the impact of algorithm initialization on the outcome, NEXI was fit to the average signal in cortex and hippocampus (separately) using N=100 random initializations covering the entire range of physical values for each parameter ($D_{i,\parallel}, D_e \in [0,3]$, $f \in [0,1]$, $t_{ex} \in [0,100]$). Based on this outcome, the range of NLLS fit initializations was further reduced to ($D_{i,\parallel} \in [1.5,3]$, $D_e \in [0.5,1.5]$, $f \in [0.1, 0.9]$, $t_{ex} \in [5, 60]$) to limit the impact of spurious noise-driven minima in voxel-wise fits.

The agreement between membrane permeability estimates derived from the experimental $t_{ex}$ values and existing literature for physiologically relevant membrane permeability values in healthy cells were compared.

### 2.5. Histology

A fixed brain sample from a 6 month-old rat was cut into 30 µm-thick slices using a cryomicrotome, and positioned on glass slides. Then, immunohistochemical stainings were performed to label various microstructure features. After a step of blocking non-specific antigens with donkey serum 5% buffer, with detergent (Triton, Sigma Aldrich, X-100, 1% 2 h incubation), a quadruple staining was prepared. It included labeling for microglial cells (anti-Iba 1, AbCam ab5076, 1/500 dilution), astrocytes (anti-GFAP, AbCam, ab7260, 1/500 dilution), neuron microfilaments (anti-NF, AbCam ab4680, 1/2000 dilution) and neuron nuclei (anti-NeuN, Millipore, MAB377X Alexa488, 1/100 dilution). Each antibody was incubated for one hour, followed by two steps of washing with PBS. Secondary antibodies were incubated at the same time, with anti-NeuN already coupled with Alexa488. We used donkey anti-chicken Cy5 (Millipore, AP194C), donkey anti-rabbit Alexa350 (ThermoFisher, 1710039) and donkey anti-goat Alexa647 (AbCam, ab150135).

Slices were mounted with Permafluor (ThermoFisher, TA-030-Fr), then fluorescence microscopy images acquired with an Axio Vision Observer microscope at x20 magnification (Carl Zeiss).

The patterns of staining intensity across the brain (mainly cortex and hippocampus) were compared to patterns of NEXI model parameters, in particular neurite density $f$.

### 3. Results
### 3.1. Impact of exchange, structural disorder and soma

#### 3.1.1. Time-dependent Standard Model parameters in GM and WM

To underline the limits of applicability of the Standard Model, Eqs (1)-(2), we evaluated SM parameter estimates in GM vs WM ROIs at different diffusion times and b-value ranges.

**Geometric parameters $f$ and $c_2$.** The apparent intra-neurite water fraction $f$ decreased with increasing diffusion time irrespective of $b_{max}$ in both GM (**Figure 2**) and WM (**Figure S1**). The decrease had a slow rate of $(2 - 3) \cdot 10^{-3}$ ms$^{-1}$ for all ROIs (**Table S1**), which would translate into an underestimation of the fraction by 0.2 points for a diffusion time of 100 ms. Neurite alignment $c_2$ increased with diffusion time, most markedly in GM.

**Compartment diffusivities.** The trends for compartment diffusivities as a function of diffusion time and $b_{max}$ were more complex. Significant decrease in parallel diffusivities (both intra- $D_{i,\parallel}$ and extra-neurite $D_{e,\parallel}$) with longer times were found in GM, but much more markedly beyond the second-order cumulant expansion regime ($b_{max} \geq 6$ (**Figure 2**, **Table S1**). The extra-axonal radial diffusivity $D_{e,\perp}$ increased with longer times only for $b_{max} \geq 6$. Time-dependence of compartment diffusivities was less pronounced in WM ROIs than GM ROIs, with only cingulum showing a reliable trend (**Figure S1**, **Table S1**).

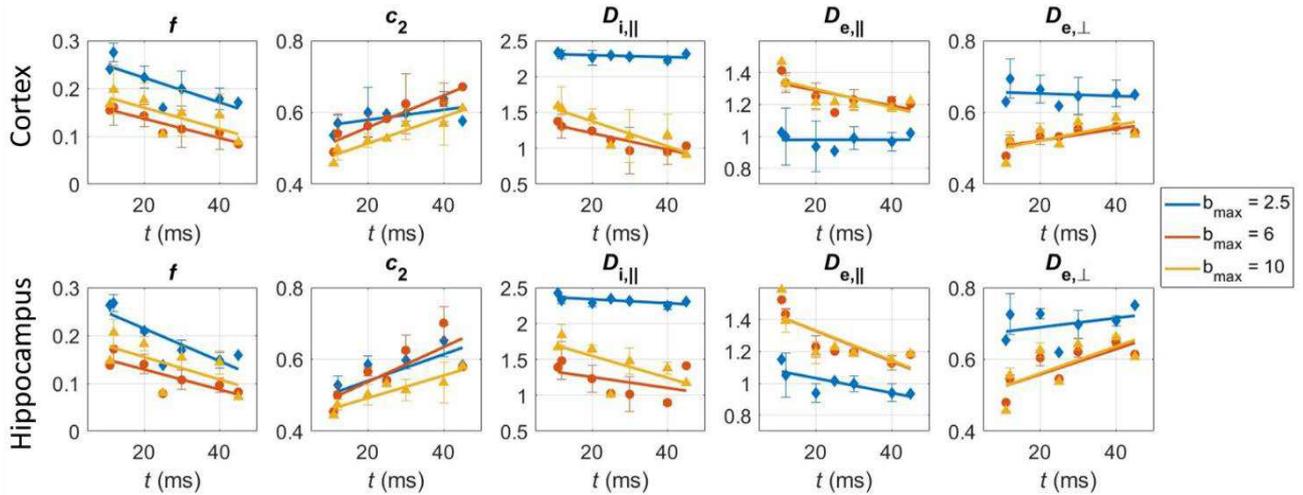

**Figure 2.** Time-dependence of SM parameters, also as a function of maximum $b$-value available, in two GM ROIs: cortex and hippocampus. *Symbols: mean ± std across rats. Solid line: linear fits.*

### 3.1.2. Exchange vs structural disorder: Time-dependent diffusion and kurtosis

As all SM parameter estimates significantly depended on diffusion time in GM ROIs, we explored whether inter-compartment exchange and/or intra-compartment non-Gaussian diffusion were relevant mechanisms in GM. To this end, we examined the time-dependence of mean diffusivity and kurtosis.

All tensor estimates were consistent across animals and displayed a reproducible ordering of rat brain structures from most coherent to least coherent: internal capsule had highest anisotropy and kurtosis, followed by corpus callosum, cingulum, cortex and finally hippocampus.

No significant time-dependence of diffusivities could be measured over the 10 – 45 ms range, based on slopes of linear fits (**Table S2**). Fitting the generic power-law formula (Eq (9) with the exponent $\vartheta$ as a free parameter) to MD($t$) yielded unreliable estimates with the exception of $D_\infty$ (**Figure 3**A for GM and **Figure S2** for WM).

Mean kurtosis on the other hand showed marked time dependence over the 10 – 45 ms range (**Figure 3** for GM and **Figure S2** for WM). The kurtosis decay in GM was close to $t^{-1}$, although the uncertainty encompasses $t^{-1/2}$ at the edge of the interval. The important observation is that the exponents of decay of MD(t) and MK(t) at long times are very different from each other, with MD at a plateau and MK still decreasing markedly. While MD(t) may still exhibit some time-dependence due to structural disorder at short times (t<20ms), this effect is practically fully coarse-grained at t>20ms such that each compartment can be approximated as Gaussian. In contrast, MK(t) has sustained sources of time-dependence throughout the 10 – 45 ms interval suggesting an additional mechanism: inter-compartment exchange. We note however that the direct comparison of the two possible power laws governing MK(t) in the GM, between exchange (~$1/t$) and 1D structural disorder (~$1/\sqrt{t}$), was not conclusive as the long-time limit was likely not reached at the longest diffusion time (45 ms); e.g. in the Kärger model the sub-leading negative $1/t^2$ term in Eq (7) still weighed in significantly (with a numerical value of 0.2 – 0.3 vs 0.5 – 0.7 for the leading term, at $t$=45ms and assuming $t_{ex}$~20 ms), which may explain the curvature for KM kurtosis in **Figure 3**C. In that regard, the Kärger kurtosis enforcing $K_\infty = 0$ was the functional form that captured best the decay of $K(t)$ at the longest diffusion times available (leftmost on the plots, **Figure 3C-D**).

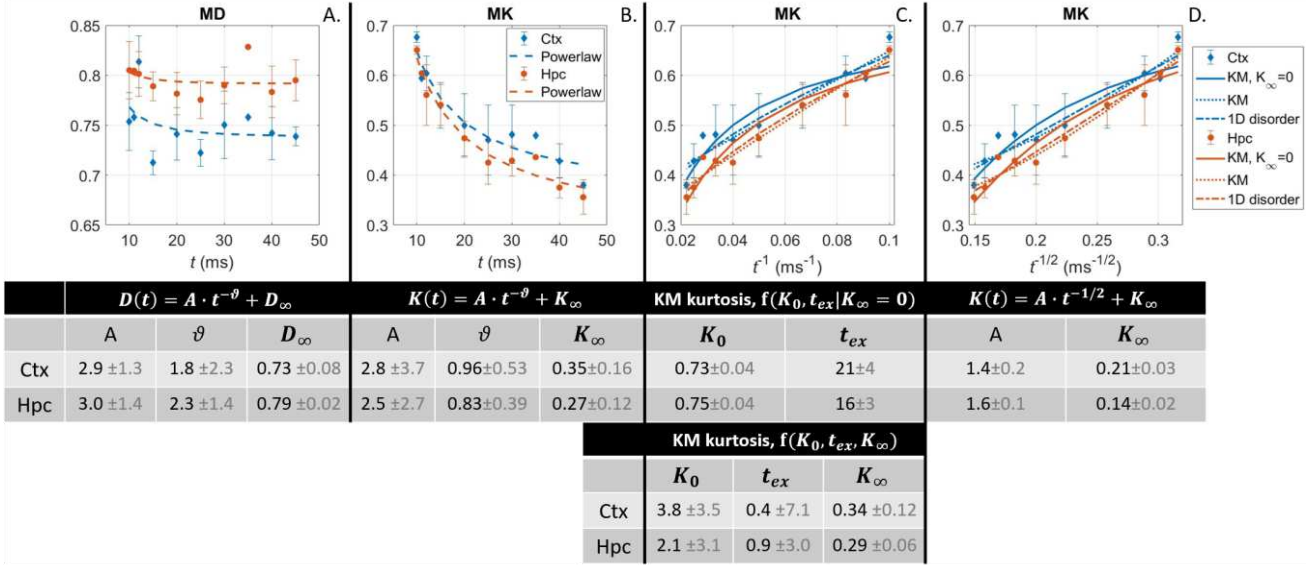

**Figure 3.** Mean diffusivity and kurtosis as a function of diffusion time, in the cortex (Ctx) and hippocampus (Hpc), averaged across animals. Fit parameters (mean±std) for each functional form are collected in the tables. A: Fitting the power-law to MD yielded very large exponent $\vartheta$ (with high variability), mainly driven by the diffusion times 10 – 20 ms. B: The behavior of MK was markedly different, with a decay throughout the 10 – 45 ms span. As a result, the power-law fit to MK yielded exponent $\vartheta$ close to 1 (and with reduced variability). C-D: The direct fitting to either the KM kurtosis (imposing $K_\infty = 0$) or the 1D structural disorder form ($\vartheta = 1/2$) showed both approaches fit the data similarly, though KM kurtosis captures the curve at the longest times (leftmost of x-axis) better. Releasing $K_\infty = 0$ in the KM results in a similar curve to 1D disorder but with poorer parameter estimates (3 free parameters instead of 2).

Allowing for nonzero $K_\infty$ when fitting the NEXI kurtosis (Eq. 7) to the measured $K(t)$ yielded extremely large uncertainty on both $K_0$ and $t_{ex}$ estimates – up to 1800% – obscuring their interpretation completely (**Figure 3**). A finite $K_\infty$~0.3 is associated with very short exchange time estimates (1 – 3 ms) which (i) are too short to be reliably estimated from our diffusion time range and (ii) correspond to a timescale where different mechanisms may also come into play, such as structural disorder.

Setting $K_\infty = 0$ enabled a more robust fit. The $K_\infty = 0$ approximation is justified in the context of the reasonable picture of a fully mixed (Gaussian) medium in a rat GM voxel at infinitely long diffusion times, due to intra-/extracellular exchange and a fully connected extracellular space. The GM yielded exchange times in the same range as the diffusion times explored and thus with better precision, e.g., $t_{ex} = 21 \pm 4$ ms in cortex and $t_{ex} = 16 \pm 3$ ms in hippocampus. Estimated exchange times were longer ($t_{ex} > 80$ ms) in the internal capsule and the corpus callosum but were also associated with a fairly large uncertainty (~35%) likely related to the inappropriate diffusion time range (10 – 45 ms) to estimate long exchange times. The cingulum (WM) displayed an intermediate behavior between GM and WM, with $t_{ex} = 43 \pm 24$ ms. This suggests that the myelin sheath plays a significant role in slowing down inter-compartment water exchange – the IC and CC are most myelinated, the GM the least, while CG may be affected by partial volume effects with neighboring GM due to its thinner structure.

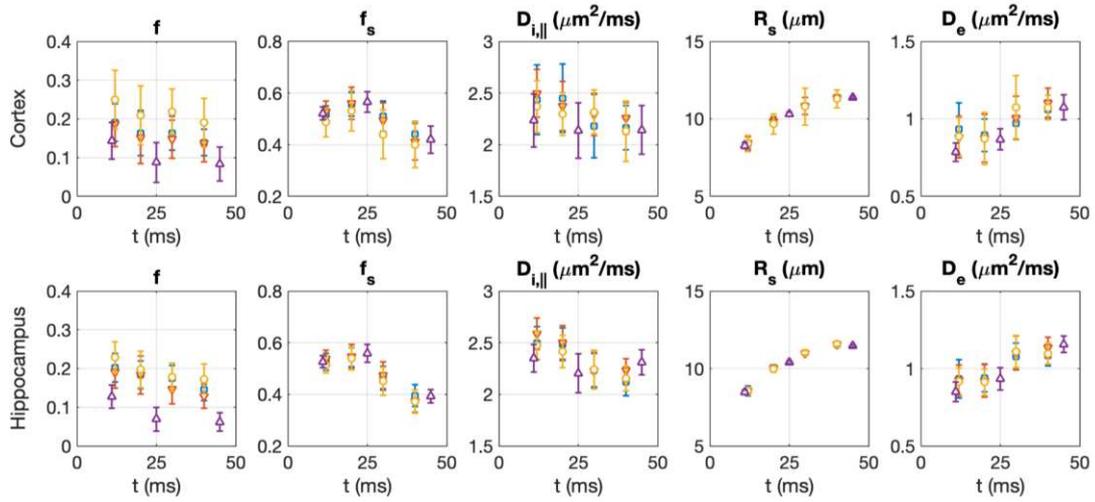

**Figure 4.** Time-dependence of SANDI model parameters for the ROI in the cortex and hippocampus. Open symbols: mean value; error bars: standard deviation over all the voxels within the ROI for each investigated rat. Note that data from Rat #1-3 were acquired at higher resolution (blue, red, yellow – voxel size 0.2x0.2x0.5 mm$^3$) and different diffusion times than Rat #4 (purple – voxel size 0.25x0.25x0.8 mm$^3$). See Table 1 for further details on the acquisition.

### 3.1.3. Exchange vs soma

To explore whether soma and/or exchange are relevant features to explain the diffusion signal in GM, we assessed whether the estimated SANDI model parameters show any significant time dependence, and then compared the quality of fit and predictions of the SANDI and NEXI models at high b-values.

The SANDI model was applied to GM ROIs at different diffusion times (**Figure 4**). We overall observed statistically significant time dependence of all SANDI model parameters, based on slopes of linear fits (**Table S3**). However, for all parameters except $R_s$ and $f$, the mean absolute percentage differences of the values at each time point with respect to the first time point at $t$ = 12 ms were within 10% at t=20 ms, suggesting that estimates of those SANDI parameters are stable for t ≤ 20 ms (see **Figure S3**). In particular, for $D_{i,\|}$ and $D_e$, this variability drops further down to within ±5% for t ≤ 20 ms.
A one-way ANOVA analysis with Bonferroni correction for multiple comparison further showed that the estimates of all SANDI model parameters (except $R_s$) were not statistically different between $t$=12 and 20 ms (see **Figure S4**). In contrast, $R_s$ estimates showed significant increase with increasing diffusion times. This is to some extent expected: $R_s$ is an MR apparent estimate of the sphere radius, weighted by the tail of the distribution, and such weighting depends on the pulse timings (Alexander et al., 2010). Therefore, our findings suggest limited bias due to exchange at $t$<~20 ms for all SANDI model parameters, except $R_s$, in vivo in rat.

Accounting for either a soma compartment or exchange between neurites and the extracellular space captures well the curvature of the signal decay as function of $b^{-\frac{1}{2}}$, which distinguishes the diffusion behavior in gray matter from that in white matter, where $\bar{S}_{|b\to\infty} \propto b^{-\frac{1}{2}}$ (an asymptotically straight line). The quality of the fit for SANDI, NEXI and other signal approximations at a single diffusion time is shown in an example dataset (**Figure 5**).

Remarkably though, a SANDI fit at short diffusion time predicted a qualitative trend of higher diffusion signals at longer diffusion times, while the experimental trend was the opposite. In this respect, the NEXI model of exchange explained signal decay curves for multiple diffusion times better than SANDI (**Figure 6**).

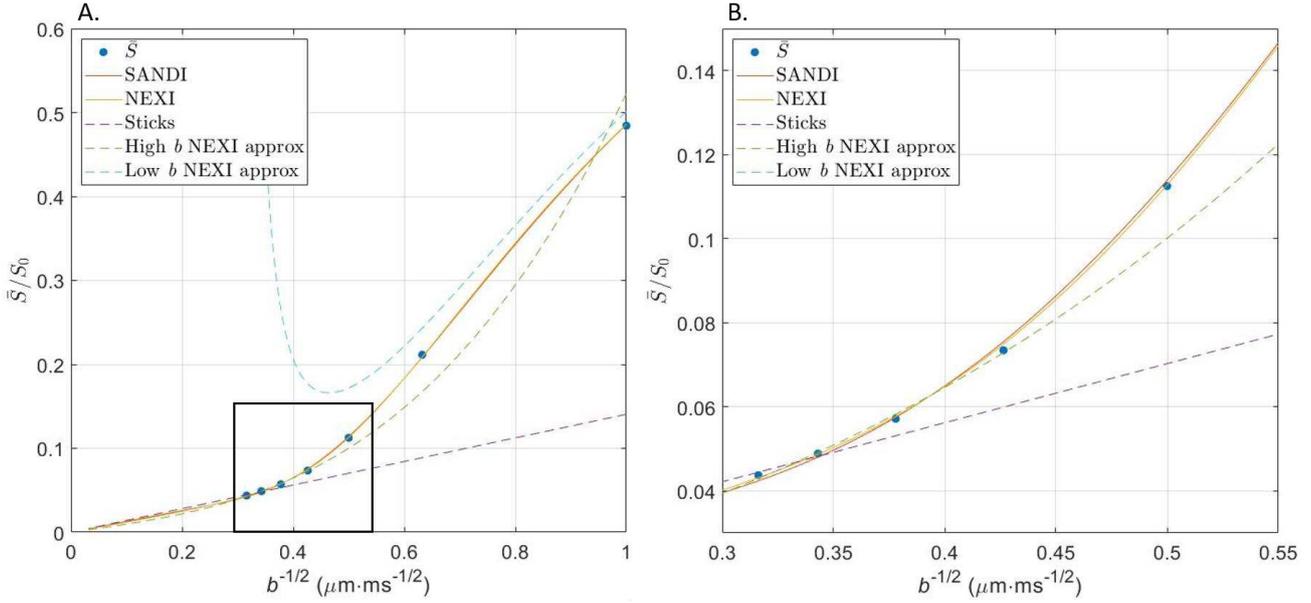

**Figure 5. A.** Various models were fit to the average signal in the cortex (Rat #2) at $t$=12 ms: SANDI, Eq (10), and NEXI, Eq (6), covering the full $b$-value range; the impermeable stick approximation (Callaghan's model); NEXI approximation at high $b$, Eq (8); and the NEXI-derived diffusivity + kurtosis approximation at low $b$ (Appendix). **B.** Zoom-in of the black framed region in panel **A**. Both SANDI and NEXI explain the data at a single diffusion time well. Callaghan's model does not describe diffusion signal decay in the cortex appropriately due to the signal's notable curvature with respect to $b^{-1/2}$, cf Eq (8). The NEXI low-b and high-b approximations are reasonable in their respective regimes. It should be noted the low-b approximation is derived from NEXI parameter estimates obtained over the entire b-value range available hence some mismatch with the experimental datapoints. The mismatch is reduced for longer diffusion times, where the Gaussian compartment approximation may be more suitable (**Figure S9**). Estimated model parameters, underlying the plotted curves: SANDI: $f = 0.22$; $f_s = 0.41$; $D_{i,\parallel} = 2.3$; $R_s = 9.3$; $D_e = 0.54$; NEXI: $f = 0.35$; $D_{i,\parallel} = 3$; $D_e = 0.73$; $t_{ex} = 20$; Sticks: $f = 0.25$; $D_{i,\parallel} = 2.4$; High b NEXI approx.: $f = 0.29$; $D_{i,\parallel} = 1.9$; $D_e = 0.35$; $t_{ex} = 12$.

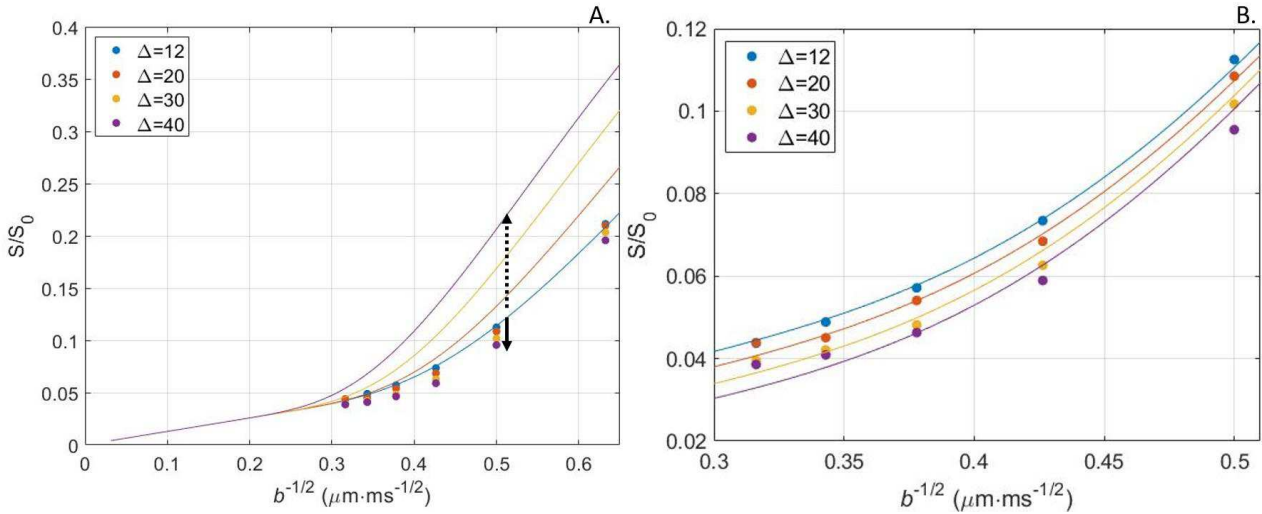

**Figure 6. A.** The SANDI model was fit to the average signal in the cortex (Rat #2) at $t$=12 ms. Estimated model parameters [$f = 0.22$; $f_s = 0.41$; $D_{i,\parallel} = 2.3$; $R_s = 9.3$; $D_e = 0.54$] were used to predict the signal for longer diffusion times (solid lines), as suggested by (Olesen et al., 2021). Qualitatively, SANDI predicted higher signal at longer diffusion times, which was opposite to the experimental pattern of increasingly reduced signal with longer diffusion time (dots). **B.** The NEXI model of exchange was fit to data from all diffusion times jointly (solid lines). The estimated model parameters were [$f = 0.29$; $D_{i,\parallel} = 2.5$; $D_e = 0.74$; $t_{ex} = 44$]. This model explained decay curves at different diffusion times well, though the agreement was poorer at the highest b-values, potentially due to an imperfect correction for Rician noise floor or to soma. *All units in μm, ms and μm²/ms*.

### 3.2. NEXI parameter estimation

Since inter-compartment exchange appears to be a relevant mechanism to explain the diffusion signal both in the low-order approximation and at high b-values, we assess the performance of NEXI – a biophysical model of two compartments with exchange – in terms of accuracy and precision in simulations, as well as feasibility and sensibility of estimated microstructure parameters in experimental rat data in vivo.

#### 3.2.1. Simulations

We first present the performance of NEXI in simulations, using Eqs (3)-(6) to generate ground truth signal, and estimating the four model parameters using conventional NLLS (see Supplementary Material for DL-based fitting, **Figures S5** and **S6**).

Scenario a: Fitting the four model parameters for each diffusion time separately. The precision was good on $D_e$ and acceptable on $f$. However, in a finite SNR case, $D_{i,\parallel}$ and $t_{ex}$ could not be estimated, irrespective of the diffusion time (**Figure 7**).

Scenario b: Fitting the four model parameters using all diffusion times jointly. In all cases, this approach significantly improved the precision on $f$ and $t_{ex}$ compared to Scenario a. Some sensitivity to $D_{i,\parallel}$ was also restored (**Figure 8**).

Varying $b_{max}$ showed that b-values larger than 2.5 ms/µm² are needed for accuracy (**Figure S7**). Both accuracy and precision were further improved for $b_{max}$ = 10 vs 6 ms/µm² but the benefits were less substantial. When inspecting the impact of $b_{max}$ given a constant number of shells, the performance of $b_{max}$ = 10 was still superior to that of $b_{max}$ = 2.5 in terms of accuracy and precision, confirming it is the b-value range that is critical for sensitivity to model parameters.

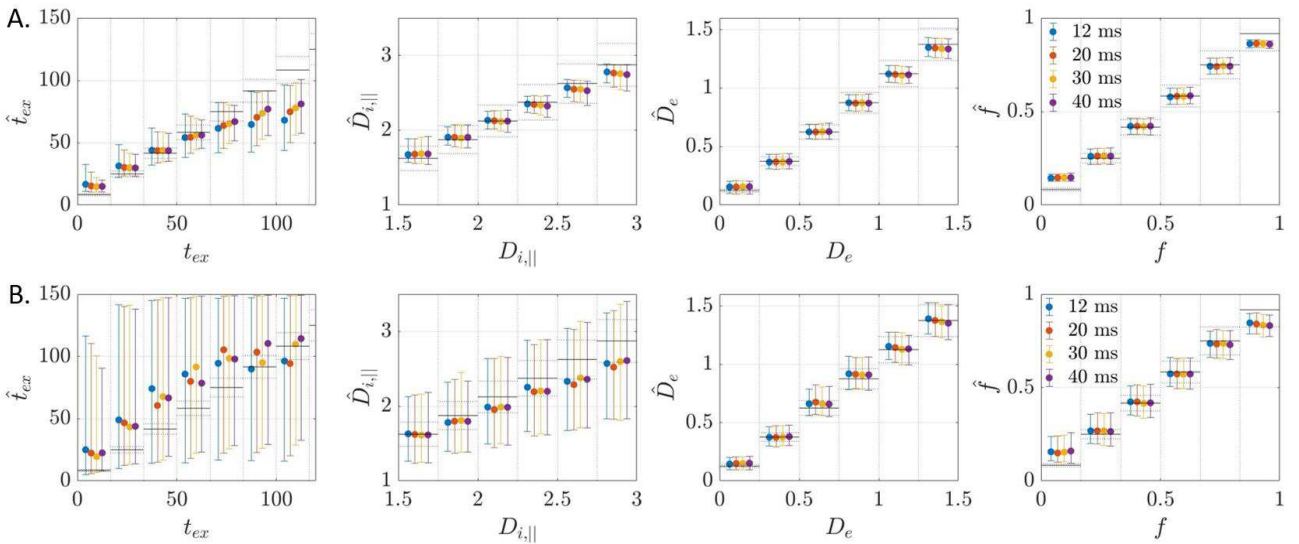

**Figure 7.** Simulation results fitting multi-shell data for each diffusion time separately using NLLS, without noise (A) or with SNR = 100 (B). Displayed is the ground truth (GT) vs estimation for 10⁴ set of random parameters. Markers correspond to the median & IQR in the corresponding intervals. Black lines are the ideal estimation ±10% error. In all cases, the precision is good on $D_e$ and acceptable on $f$. However, in a finite SNR case, $D_{i,\parallel}$ and $t_{ex}$ cannot be estimated, irrespective of the diffusion time.

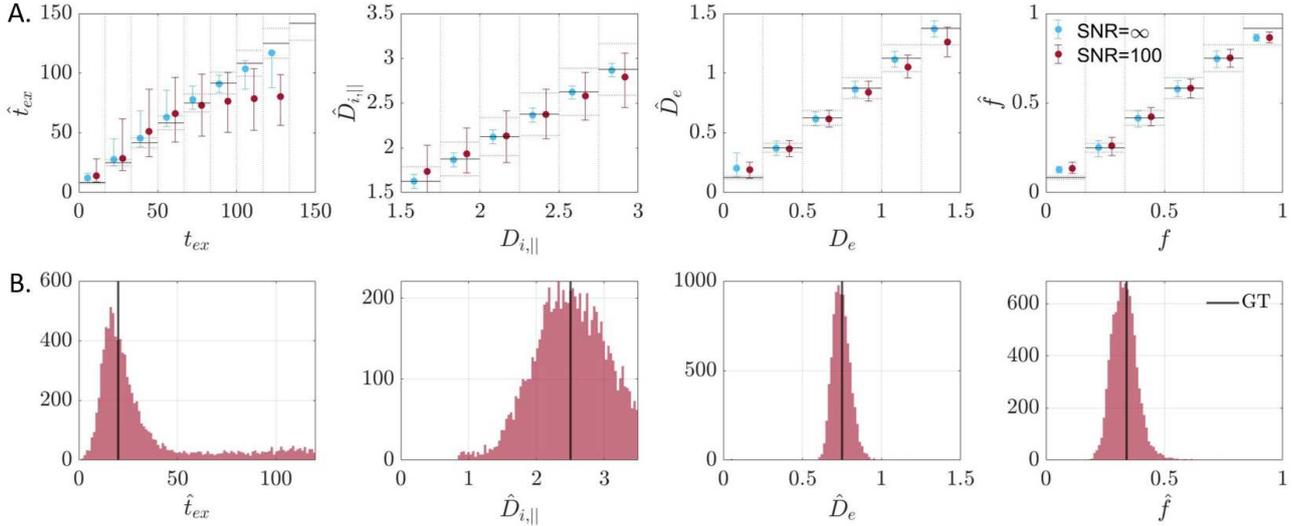

**Figure 8.** Simulation results fitting multi-shell multi-$t_d$ data jointly using NLLS, for random GT (A) or fixed to $[t_{ex}^{th}, D_{i,\parallel}^{th}, D_e^{th}, f^{th}] = [20, 2.5, 0.75, 0.34]$ (B). **A**: Displayed are the medians & IQR in each bin. Black lines: ideal estimation ±10 % error. Without noise, NLLS fits all parameters with high accuracy and precision. At SNR=100, uncertainty increases primarily for $D_{i,\parallel}$ and $t_{ex}$ and sensitivity to high $t_{ex}$ values is lost but the performance is much improved compared to single $t_d$ fits (Fig. 7). **B**: At SNR=100, good accuracy is achieved for all NEXI parameters. For $D_{i,\parallel}$ the precision is poor. *Black solid line: ground truth*.

### 3.2.2. Experimental: in vivo rat GM

The time-dependence analysis of diffusion, kurtosis and SM metrics highlighted the sharp difference in behavior between highly myelinated white matter fibers such as the internal capsule and the corpus callosum, and GM. Results from the previous sections suggest non-negligible inter-compartment water exchange in GM, which should be accounted for by biophysical models of this type of tissue when working at relatively long diffusion times (*t*>20 ms).

Simulation results for NEXI performance in turn suggested the use of multi-shell multi-*t* data was crucial for the reliable estimation of model parameters, and of the exchange time in particular, for a broad range of ground truth values.

Given a dataset comprised of *b*-values up to 10 ms/μm² and three to four diffusion times, all NEXI model parameters could be estimated both at the ROI and at the single voxel level in the rat GM.

Testing 100 random initializations on ROI-averaged signal showed that the overwhelming mode of the distribution of outcomes corresponded to a solution where $D_{i,\parallel} > 2$ and $D_e < 1$ (**Figure S8**).

Parametric maps of NEXI estimates were consistent with expected neuroanatomy of the rat brain (**Figure 9A-D**). Simulations predicted the variability was largest for $t_{ex}$ and $D_{i,\parallel}$. ROI-based analysis in the cortex and hippocampus confirmed these trends and also revealed good between-subject consistency (**Figure 9E**). On average, the intra-neurite diffusivity was 2.5 μm²/ms, the extra-neurite diffusivity was 0.75 μm²/ms and the neurite fraction was around 0.3.

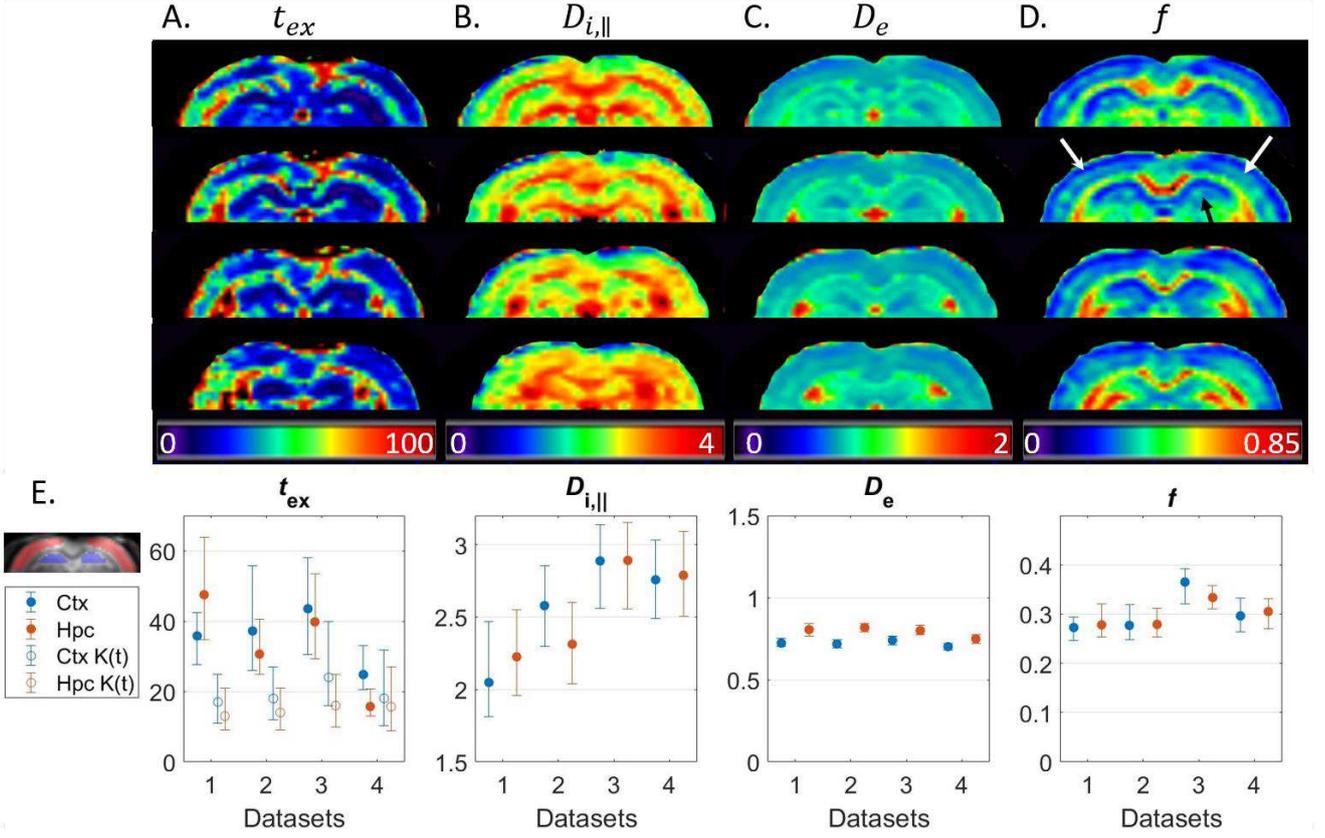

**Figure 9. A-D:** Four coronal slices of NEXI parametric maps calculated using NLLS from a multi-shell multi-$t$ dataset. The maps enable a good differentiation between GM & WM as well as between different cortical layers (white arrows) or hippocampal subfields (black arrow). **E**: Median & IQR of model parameters in the cortex and hippocampus ROIs across the four datasets. The exchange time estimate is also compared with $t_{ex}^{K(t)}$, Eq (7). Experimental trends agree with the simulations. Regarding $t_{ex}^{K(t)}$, the estimation agrees with $t_{ex}$ very well for Dataset #4, which had the highest SNR (larger voxels), and is otherwise shorter.

For Datasets 1 – 3, the estimated exchange time $t_{ex}$ was higher with NEXI than with $t_{ex}^{K(t)}$: 30 – 60 ms vs 10 – 40 ms, respectively, in cortex and 25 – 65 ms vs 10 – 25 ms, respectively, in hippocampus. Remarkably, for Dataset 4 (x2.5 voxel volume compared to Datasets 1 – 3), the two approaches displayed better agreement and estimates for $t_{ex}$ ranged 10 – 30 ms.

The exchange time can be related to the cell membrane permeability of a cylinder via $P = \dfrac{d}{4\left(t_{ex} - \dfrac{d^2}{32\,D_{i,\parallel}}\right)}$, where d is the diameter of the cylinder (neurite) and P is the *diffusional water membrane permeability* (affected by the properties of the lipids in the membrane and by water-channel proteins embedded in the membrane and different from the osmotic permeability, generally larger and measured in the presence of an osmotic pressure gradient over the membrane) (Meier et al., 2003; Markus Nilsson et al., 2013). Given the typical diameter and diffusivity values, this is further very well approximated as $P \cong \dfrac{d}{4 t_{ex}}$. Assuming d ~[0.5 - 2] µm, a characteristic exchange time $t_{ex}$= [15 – 60] ms, as estimated here using NEXI, yields P ~ [2.1 – 33] x $10^{-3}$ µm/ms.

The intra-neurite fraction map displayed substantially larger values in white matter than gray matter though the model does not in principle support white matter (the extra-neurite space cannot be assumed to be isotropic). This can be explained by the fact that myelin is MR-invisible in our diffusion MRI measurements (due to the long TE) and the physical space occupied by myelin is therefore not considered. Assuming neurite physical occupancy fractions are similar in GM and WM, the *relative* neurite fraction we estimate is higher in WM because the myelin space reduces the extracellular space. Different $T_2$ relaxation times in compartments between GM and WM could also account for the difference. The other NEXI parameters were also higher in WM, which is in good agreement with more aligned structures (enabling faster diffusivity) and longer exchange times in highly myelinated axons.

All NEXI parameters also showed contrast within GM structures, such as across cortical layers and hippocampal subfields. Notably, the neurite fraction was higher in central cortical layers, consistent with neurofilament staining in ex vivo rat cortex slices (**Figure 10**). A higher NEXI neurite fraction in the central section of the hippocampus (dorsal dendate gyrus) agreed especially with higher astrocyte density in that region, which suggests astrocytic processes also contribute to this parameter via their similar geometry to neurites, i.e. long, thin structures.

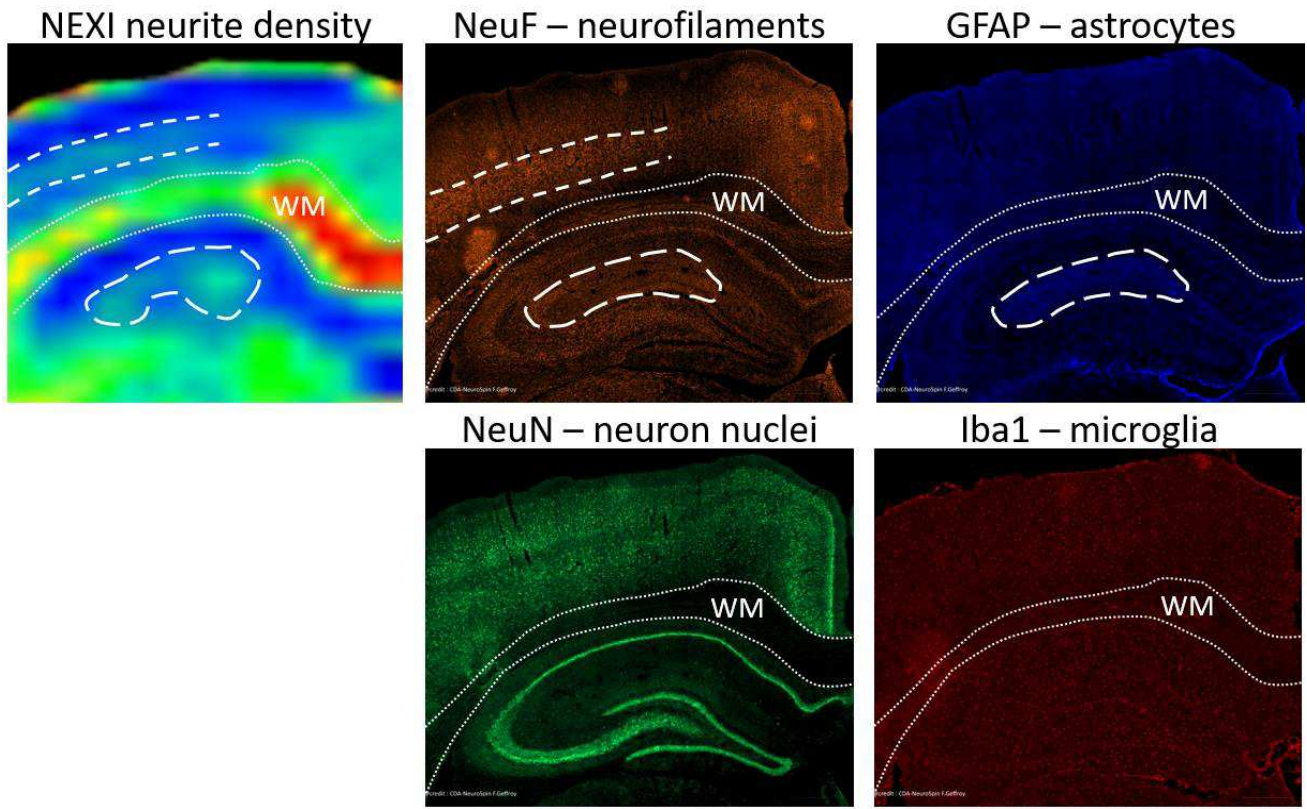

**Figure 10.** Features of NEXI neurite density map features as compared to cellular components obtained from histological stainings: neurofilaments (orange), astrocytes (blue), neuron nuclei (green) and microglia (red). The WM is outlined in fine dotted lines for legibility; cortex lies above, hippocampus below. Higher NEXI neurite density in central cortical layers agrees with higher density of neurofilament staining (dashed lines). Higher NEXI neurite density in the central part of the hippocampus (dorsal dendate gyrus) agrees especially with higher density of astrocytes but also neurofilaments (long-dashed contour). Neuron soma and microglia do not seem to contribute to NEXI neurite density contrast.

## 4. Discussion

In this work, we propose NEXI as an extension of the SM of diffusion suitable for GM. NEXI accounts for inter-compartment exchange between neurites and the extracellular space, building on the anisotropic Kärger model of two exchanging compartments. Using multi-shell multi-*t* datasets acquired in the rat brain *in vivo*, we investigate the suitability of NEXI to describe diffusion in the GM, compared to other approaches such as SM, structural disorder, or the addition of a soma compartment. We identify exchange as the mechanism that best explains diffusion-time-dependence of signal in both low-*b* and high-*b* regime, and thereby propose NEXI as the minimal model for GM microstructure mapping. We finally propose multi-*b* multi-*t* acquisitions schemes as best suited to estimate NEXI model parameters $[f, D_{i,\parallel}, D_e, t_{ex}]$ reliably.

**SM applicability.** The presence of exchange yields a spurious time-dependence of SM parameters. Unsurprisingly, the time-dependence of geometric parameters for the SM was most marked in GM and cingulum. The apparent intra-neurite fraction decreased with increasing diffusion time. Qualitatively, this can be interpreted as water molecules that leave the intra-neurite space developing a diffusion signature closer to hindered diffusion – as the extracellular space or large soma – rather than restricted and unidirectional along the neurite. The neurite alignment also increased with longer times. These results suggest that in the SM the diffusion time acts as a filter that attributes to the intra-neurite space only cellular processes (axons, dendrites and glial processes) that can be considered impermeable over that time scale. Arguably, only the more myelinated and aligned neurites are retained at longer times.

The time-dependence of SM compartment apparent diffusivities varied with *b*-value regime, with more pronounced trends for $b_{max} \geq 6$ ms/μm$^2$ and for GM than for WM ROIs. A trend of apparent diffusivity decreasing with time can be a signature of either non-Gaussian diffusion, exchange with a slower compartment or both. A trend of apparent diffusivity increasing with time is compatible with either an exchange-dominated regime where the exchange happens with a faster-diffusing compartment or with a gradual decrease of compartment anisotropy as the molecules explore a larger space (e.g. $D_{e,\parallel}$ and $D_{e,\perp}$ are becoming more similar at longer times in cortex and hippocampus). Essentially, in GM the SM parameters become ill-defined and their interpretation in terms of microstructure becomes challenging.

Our analysis confirms that the SM is applicable in thick WM bundles. The quantitative estimate of compartment diffusivities may however depend on the *b*-value range (the anisotropy of the extra-neurite compartment in particular may be exacerbated) and the intra-neurite fraction may decrease with longer times likely by dropping unmyelinated axons in WM. The latter are more numerous in the rodent than in the human brain (Wang et al., 2008) so this effect may not impact human brain estimates as much. However, the SM assumptions are likely not met in GM and this could also impact WM bundles that have substantial partial volume with cortex, such as the cingulum. In particular, while in the cumulant regime ($b \lesssim 2.5$) only a progressive filtering of neurites with increasing time was observed, for higher *b*-values apparent compartment diffusivities were also affected, particularly in the intra-neurite space.

**Exchange vs structural disorder in cortex.** Our data show negligible time-dependence of mean diffusivity, over the range of diffusion times 20 – 45 ms, in the rat GM. This is consistent with previous reports for the *in vivo* rat brain by (Pyatigorskaya et al., 2014) and *ex vivo* mouse brain by (Aggarwal et al., 2020). We note however that pronounced OGSE frequency dependence of diffusivities has been reported at shorter time scales in rat and mouse cortex (Aggarwal et al., 2020; Does et al., 2003; Pyatigorskaya et al., 2014), and, in the case of the work by Does et al., has been attributed to the neurites based on the exponent $\vartheta = 1/2$ by (Novikov et al., 2014).

Weak diffusivity time-dependence has also been highlighted in human cortical gray matter for diffusion times 21 – 100 ms (Lee et al., 2020b). The human brain presents the additional challenge of thin cortical ribbons and thereby relatively strong partial volume with white matter and CSF in "cortical voxels". The rat brain is, from this perspective, a well-suited model system for exploring cortical properties. We note however the weak time-dependence of diffusion is not necessarily transposable to other organs, *ex vivo* conditions and different diffusion time ranges. For example, (Jespersen et al., 2018) have shown significant time-dependence of the diffusion coefficient in fixed pig spinal cord, for diffusion times 6 – 350 ms. The negligible time-dependence of *D* in our context suggests/validates that, within the cumulant expansion regime, the tissue can be considered as a collection of Gaussian compartments, one of the main assumptions behind multi-compartment models of diffusion, and that structural disorder is therefore negligible at our diffusion time scales. A larger dynamic range of diffusion times, covering at least a logarithmic decade, may be necessary to detect diffusivity time-dependence.

On the other hand, kurtosis displayed marked time-dependence, also in agreement with findings of all afore-mentioned studies (Aggarwal et al., 2020; Jespersen et al., 2018; Lee et al., 2020b; Pyatigorskaya et al., 2014). Together with the absence of marked time-dependence of *D*, the decrease in $K(t)$ with $t$ can be attributed to inter-compartment exchange, Eq (7) (Fieremans et al., 2010; Jensen et al., 2005; Kärger, 1985) rather than structural disorder, with kurtosis decaying to zero at very long times when compartments are fully mixed and appear as a single Gaussian compartment. This interpretation was further supported by a power-law fit to the measured $K(t)$ compatible with a decay as $1/t$ for mean kurtosis, though the long-time limit may not have been reached, which challenges the estimation of this exponent. Furthermore, in the intermediate diffusion time regime explored, the Kärger kurtosis captured the trend of the longest diffusion times better than $1/\sqrt{t}$ functional form, in contradiction to (Lee et al., 2020b). The exploration of a broader range of diffusion times in future work may enable a more definite assessment of the most relevant power law of $K(t)$ decay and of the relative contribution of the competing effects of incomplete coarse-graining over the structural disorder, and inter-compartmental exchange.

It should be underlined that the functional form for structural disorder as $1/\sqrt{t}$ corresponds to one-dimensional short-range disorder which is potentially suited for intra-neurite diffusion (Novikov et al., 2014). In principle, structural disorder could also arise from extracellular water, and would in this case be expected to follow the functional form for 2d or 3d disorder, as $(\ln t)/t$ or $1/t$. This functional form should however be followed by both $D(t)$ and $K(t)$. Overall, the trend in $D(t)$ was flat and certainly did not support a decay as $(\ln t)/t$ which is more pronounced than $1/\sqrt{t}$. The 2d or 3d disorder was also not supported by $D(t)$ in human cortex (Lee et al., 2020b).

The estimation of inter-compartment exchange based on NEXI $K(t)$ yielded relatively long exchange times ($t_{ex}^{K(t)}$ = 80 – 130 ms, exceeding our diffusion time range) in highly myelinated white matter bundles such as the corpus callosum and internal capsule, intermediate exchange times ($t_{ex}^{K(t)} \sim$ 40 ms) in thinner bundles such as the cingulum that may experience partial volume effects with neighboring gray matter, and relatively short exchange times in the cortex and hippocampus ($t_{\mathrm{ex}} \sim$ 15 – 20 ms).

Estimates of exchange time in WM bundles were unsurprisingly imprecise due to the mismatch between probed timescales (10 – 45 ms) and the expected exchange time $t_{ex}^{K(t)}$ > 80 ms. Nevertheless, this result validates *a posteriori* the assumption of non-exchanging compartments for white matter models at diffusion times typical for PGSE acquisitions: *t* < 80 ms. Our findings are also consistent with previous studies in the human WM reporting exchange times above 500 ms (Lampinen et al., 2017; Nedjati-Gilani et al., 2017), and 350 – 400 ms in mouse corpus callosum (Hill et al., 2021).

In the case of gray matter, the $t_{\mathrm{ex}}^{K(t)}$ estimates of 15 – 20 ms were consistent with previous studies in human gray matter (Veraart et al., 2018a) and perfused pup rat spinal cord (Williamson et al., 2019). Other studies using relaxation-based methods suggested however longer exchange times of 100 – 150 ms in astrocyte and neuron cultures (Yang et al., 2018), in rat subcortical structures – presumably the striatum (Quirk et al., 2003) and in rat perfused cortical cultures (Bai et al., 2018). Filter-exchange imaging (FEXI), another diffusion-based method to estimate the exchange time between a slow and a fast water pool reported an exchange time on the order of 1 s in WM and 2.5 s in GM using a filtering block of $b_f$ = 0.9 ms/μm$^2$ (Lampinen et al., 2017; M. Nilsson et al., 2013). While the exchange times between different WM tracts agreed with expected myelination levels, e.g. up to 3 s in corpus callosum and 500 ms in anterior corona radiata, it is somewhat counterintuitive that the exchange time would be longest in GM – the authors suggested the latter was likely overestimated. Recent work using FEXI with a similar filter also yielded an exchange time of around 1 s in WM and 1.4 s in GM (Bai et al., 2020), while arguing that there is no direct evidence that what FEXI measures is the exchange between intra- and extra-cellular compartments.

Indeed, while bi-exponential functions typically fit diffusion decay in brain tissue well, the association of the slow and fast water pools to specific tissue compartments has never been straightforward (Kiselev and Il'yasov, 2007; Novikov et al., 2018a), in particular since a distribution of non-parallel sticks – a single compartment, technically – also yields a characteristic decay that is well approximated by a bi-exponential function (Assaf and Cohen, 1998; Callaghan et al., 1979; Novikov et al., 2018a; Sehy et al., 2002). For this same reason, alternative approaches to the Kärger model, extracting the exchange time from the decay of the intra-cellular fraction – as estimated from a bi-exponential model – with increasing diffusion time (Moutal et al., 2018) were explored in yeast suspensions but are not suitable for brain tissue. Indeed, predominant stick-like geometries in both white and gray matter invalidate the approximation of a sum of two Gaussian compartments in any direction or for the powder-average signal.

Our analysis of $K(t)$ so far cannot provide direct information on the mechanisms of exchange, such as intra/extracellular, neurite/soma or neurite/neurite exchange. However, numerical simulations suggest that neurite/soma and neurite/neurite exchange within the same neuron occur at longer time scales (on the order of 100 ms or more) than those estimated here (~20 ms) (Ianus et al., 2020) and support the intra/extracellular exchange as the dominant mechanism in these experiments.

**Exchange vs soma**. Going a step further, we investigated the performance of a two-compartment model with exchange (NEXI) and of a three-compartment model accounting for soma (SANDI) to capture diffusion signal decay at high *b*-values, and for multiple diffusion times. SANDI extends the SM by adding an extra compartment for modelling explicitly diffusion restricted in soma and relies on the assumption of negligible exchange between the three tissue compartments: intra-neurite, intra-soma and extra-cellular. Our results do not challenge this assumption in the rat GM in vivo for relatively short diffusion times ($\leq$ 20 ms), while challenging it for longer diffusion times (> 20 ms), where the SANDI model parameters show some time-dependence. This diffusion time cutoff is in line with $t_{ex}^{K(t)}$ and suggests that unaccounted exchange mechanisms between the three major tissue compartments in GM (cellular processes, soma and extra-cellular space) may bias SANDI parameters estimation at diffusion times longer than 20 ms. On the other hand, our results also suggest that SANDI model parameter estimation provides $f$, $D_{i,\parallel}$ and $D_e$ estimates in good agreement with the equivalent counterpart from the SM and NEXI. The importance of modeling exchange in addition to soma was mostly evident in the ability of NEXI vs SANDI to predict signal decay curves for longer diffusion times based on model parameters estimated at short diffusion times. This result is consistent with findings in the rat cortex ex vivo (J. L. Olesen et al., 2021b). However, based on the NEXI parameter estimation performance alone – discussed below – a larger *q-t* coverage and higher SNR would likely be needed in vivo to account for both soma and exchange in a model.

**NEXI parameter estimation**. To provide recommendations of minimum data and fitting procedures for NEXI, we first established its performance in simulations. Given a comprehensive protocol with 7 shells up to $b_{max}$ = 10 ms/μm$^2$ and high final SNR of 100 – boosted by the MP-PCA denoising procedure and the powder-averaging over directions – data at a single diffusion time were insufficient to estimate $t_{ex}$ and $D_{i,\parallel}$. Noise was clearly the culprit as the noiseless simulations otherwise demonstrated good performance for all four model parameters. Fitting the NEXI model to joint data over four diffusion times dramatically improved the accuracy and precision for all four model parameters though $D_{i,\parallel}$ remained the most challenging parameter to estimate, consistent with other model frameworks (Jelescu et al., 2016; Novikov et al., 2018b; Palombo et al., 2020). The benefit of a broader *b*-value range was critical between 2.5 and 6 ms/µm², but only marginal beyond, which suggests a range 0 < *b* < 6 ms/µm² could be sufficient to estimate NEXI parameters.

Simulations also suggested the exchange time estimate $\hat{t}_{ex}$ plateaus beyond ground truths $t_{ex} \geq 80\ ms$ approximately. This is likely related to the diffusion time range simulated 12 – 40 ms, which is too short to probe slow processes with longer characteristic exchange times. For tissues where longer exchange times are expected, the diffusion time range should be adjusted accordingly.

Importantly, the performance of NEXI on experimental data using a multi-shell multi-*t* protocol was consistent with simulations. On average, the intra-neurite diffusivity was 2.2 – 2.5 µm²/ms, in agreement with its estimate from the SM and from SANDI at the shortest diffusion times, as well as with previous reports of intra-neurite/axonal diffusivity (Dhital et al., 2019; Kunz et al., 2018; J L Olesen et al., 2021). The extra-neurite diffusivity was 0.75 µm²/ms and remarkably also agreed with the SANDI estimate at the shortest times. We underline that the intra-neurite diffusivity corresponds to the parallel diffusivity, with $D_{i,\perp} = 0$ in the perpendicular direction (the stick picture). A three-fold ratio between $D_{i,\parallel}$ and $D_e$ is consistent with previous literature that reported similar Apparent Diffusion Coefficient (ADC) between intra- and extra-cellular water in rat GM (Duong et al., 1998). Considering the picture of isotropically-oriented neurites, the ADC of intra-cellular water in any given direction would be estimated at $D_{i,\parallel}/3$ and thereby similar to $D_e$.

The neurite fraction was about 0.3, which is lower than estimates from ex vivo histology (~0.65) but nonetheless higher than the SANDI estimate for neurite fraction at the shortest diffusion time (f ~0.25). By comparing NEXI and SANDI compartment fractions, it appears the soma is associated to extra-neurite space in NEXI. While accounting for exchange had the advantage of providing a time-independent estimate of the neurite fraction, thereby correcting for its decrease with longer times in models of non-exchanging compartments, the absolute value of the neurite fraction estimate remains lower than the 60 – 70 % expected from histology (Bondareff and Pysh, 1968; Motta et al., 2019; Shapson-Coe et al., 2021; Spocter et al., 2012). Combined relaxation-diffusion measurements may help improve the quantification of the neurite fraction by correcting for relaxation time-weighting (Barakovic et al., 2021; Hutter et al., 2018; Tax et al., 2021; Veraart et al., 2018b). Exchange processes on a shorter scale than those explored here also cannot be excluded, as very short exchange times have been recently reported in rat brain ex vivo (J. L. Olesen et al., 2021b).

Parametric maps agreed with known rat brain structure, with clear delimitation between gray and white matter. Contrast between adjacent cortical layers and between hippocampal sub-fields was also apparent. Comparison with histological staining revealed that higher NEXI neurite fraction in middle cortical layers corresponded to higher neurofilament density in that area, but that in hippocampus, abundant astrocytic processes could contribute to the higher NEXI neurite fraction. Indeed, water is ubiquitous and microstructural features with similar geometry are typically pooled together – here neurites and astrocytic processes are both thin elongated cylinders. This is expected to be the case for all water diffusion models proposed, though the balance of contributions between neurons and glial cells has never been firmly established. The exchange time estimates depended on the underlying SNR of the data and on the estimation approach. NEXI estimated an exchange time of 15 – 60 ms while the matching estimate from $K(t)$ was 10 – 40 ms. These exchange times, combined with intra-neurite realistic diameter values yielded a range of cell membrane permeability values on the order of $P \cong [2.1 - 33] \cdot 10^{-3}$ µm/ms. This range of permeability values is in agreement with previous reports of physiologically relevant membrane permeability values in healthy cells: $P \cong [6 - 30] \cdot 10^{-3}$ µm/ms (Baylis, 1988; Harkins et al., 2009; Latour et al., 1994; Stanisz et al., 1997; Vestergaard-Poulsen et al., 2007), and, as expected, lower than the permeability of red blood cells $P_{RBC} \cong [30 - 60] \cdot 10^{-3}$ μm/ms, which are known to be highly permeable and optimized for exchange (Benga et al., 2000).

The importance of high SNR (> 20 – 30 in an individual *b*=0 image) for reliable parameter estimation was manifest throughout our data which had a strong SNR spatial gradient from cortex to deep brain due to the use of a surface transceiver. An experimental setup with a volume coil for transmission and a surface coil for reception would yield uniform SNR and enable estimates of NEXI parameters over the whole brain. DL approaches are increasingly replacing NLLS in biophysical model estimation (Hill et al., 2021; Nedjati-Gilani et al., 2017; Palombo et al., 2020). We also showed here an improved management of noise by DL vs NLLS, although bias towards the mean of the prior was also more pronounced and DL outputs should always be examined carefully (Coelho et al., 2021; Gyori et al., 2021b; Martins et al., 2021). Alternatively, a log-likelihood objective function could be considered for the NLLS optimization to account for Rician noise in a voxel-wise fashion. As there was already good agreement between sum-of-squares NLLS and DL on our data, we did not implement the log-likelihood minimization, but it may be worth considering for lower SNR data.

**Limitations**

We note that a multi-shell multi-*t* acquisition protocol may be difficult to implement when scan time is of the essence. Future work will focus on optimizing the protocol to the best compromise between minimal scan time and maximal accuracy and precision of model parameter estimates. The currently large uncertainty on $t_{ex}$ and $D_{i,\parallel}$ estimates will also benefit from such an optimization, which may include schemes that combine multiple diffusion tensor encodings (Chakwizira et al., 2021).

After inspecting the mode of NEXI outcomes as a function of random algorithm initialization, and in line with recent evidence that $D_{i,\parallel} \cong 2 - 2.5$ μm²/ms (Dhital et al., 2019; Howard et al., 2020; Kunz et al., 2018; J L Olesen et al., 2021), we chose an algorithm initialization where $D_{i,\parallel} > D_e$ in NLLS, and trained the DL network on disjoint intervals $D_{i,\parallel} \geq 1.5$ and $D_e \leq 1.5$. These constraints would likely need to be reconsidered and relaxed, particularly when characterizing pathological conditions or diseases, such as Alzheimer's or Huntington's diseases, where tangles and/or proteins accumulate in the intracellular space, hence increasing the cytoplasm tortuosity and reducing $D_{i,\parallel}$, potentially to the level of being slower than $D_e$.

One substantial limitation of the NEXI model is that it does not account for soma as a third compartment, which are then artificially absorbed into the extra-neurite space as the relative fractions suggest. Recently, the evidence of curved boundaries in the cortex, attributed to soma, was presented by observing the localization regime of diffusion on a human Connectom scanner (Lee et al., 2021). An important line of future work is the extension of NEXI to three compartments, or, equivalently, the incorporation of exchange processes in the SANDI model. For now, SANDI accounts for three non-exchanging compartments – soma, neurites, extracellular space – but thus requires data acquisition at relatively short diffusion times (≤ 20 ms) concomitantly with high *b*-values ($b_{max} \geq 6$ ms/μm²) (Palombo et al., 2020) which can only be achieved on preclinical and Connectom scanners, but not on typical clinical scanners. Going forward, accounting for exchange in SANDI will make it translatable to clinical settings, by enabling the use of longer diffusion times.

Furthermore, NEXI considers Gaussian compartments, an assumption which seems to break for the neurite compartment, as revealed at higher *b*-values, likely due to finite length of dendritic processes, branching, etc. This poses a significant conundrum, as low *b*-values where non-Gaussian contributions in each compartment can be neglected are detrimental for estimation accuracy and precision, while higher *b*-values reveal non-Gaussian effects which bias the outcomes as well. Ultimately, computational models based on realistic simulations of neural cells (Callaghan et al., 2020; Ginsburger et al., 2019; Lee et al., 2020a; Palombo et al., 2019) and cortical substrate may be the best approach for characterizing gray matter microstructure *in vivo*.

Analytical approaches able to incorporate intra-compartmental non-Gaussian time-dependent effects, such as (Lee et al., 2018), are in need to faithfully quantify the structural disorder contributions.

Finally, built on the anisotropic KM, NEXI assumes exchange happening within each ensemble of "neurite + its immediate extracellular space" separately. This may differ from GM microanatomy, where neurites at different angles can be piercing a volume of the size of the diffusion length. Sequential exchange processes can bring a molecule from, e.g., one stick to the extra-stick space to a differently-oriented stick, and so on; such a model geometry has not been considered. We note, however, that in the limit of $bD_e \gg 1$, the distinction between this more general geometry and NEXI should vanish, since, once a spin enters the extra-neurite space, its contribution to the overall signal gets exponentially suppressed, as does the memory about the above sequential exchange processes with different orientations. Hence, we expect that $t_{ex}$ estimated from the full protocol including strong diffusion weightings will be more accurate than that from the time-dependent kurtosis approximation of Eq (6).

**Value**

Taken together, our results suggest that inter-compartment exchange is not negligible in gray matter at typical PGSE or clinical diffusion times ($t > 20$ ms) and should therefore be accounted for in biophysical models of gray matter and potentially even in thinner white matter tracts such as the cingulum (in rodents), and by extension to demyelinating WM as a result of disease. Our findings also highlight an additional challenge for approaches that use *b*-tensor encoding techniques to disentangle various tissue geometries or solve model degeneracy (Afzali et al., 2021; Coelho et al., 2019; Gyori et al., 2021a; Reisert et al., 2019). Since free gradient waveforms introduce by design a whole spectrum of relatively long diffusion times (> 20 ms), the ill-definition of the diffusion time may become problematic in a regime where exchange cannot be neglected.

NEXI constitutes an important first step in accounting for inter-compartment exchange in GM and developing a more realistic model of diffusion in gray matter. The estimate of the exchange time alone can be used as a proxy for membrane permeability, which is known to increase with injury or neurodegeneration (M. Nilsson et al., 2013; Pacheco et al., 2015), and could yield an original and valuable new biomarker of tissue integrity, metabolism and function (Bai et al., 2018). In myelinated structures, the exchange time could also become a strong proxy for the myelin thickness, which is at the heart of several "in vivo histology" efforts (Brusini et al., 2019; Hill et al., 2021; Lazari and Lipp, 2021; Mancini et al., 2020).

## 5. Conclusions

One fundamental challenge in brain microstructure is to establish the biophysical origin of effects beyond the "Standard Model" (SM) picture of non-exchanging Gaussian compartments. Here we showed that in the rat GM in vivo, the exchange dominates over structural disorder, and offer the picture of diffusion time effectively filtering out the contribution of unmyelinated neurites with stronger dispersion. At long times, this picture suggests that only the myelinated (non-exchanging) neurites contribute to the intra-neurite SM compartment, and the rest is asymptotically attributed to extra-neurite space. Exchange also explains signal decay curves across different diffusion times better than the addition of a soma compartment. If a choice is warranted, a two-compartment model with exchange – NEXI – is better suited than a three-compartment model with soma for characterizing cortical microstructure at diffusion times $t > 20$ ms, while also yielding a valuable estimate of exchange time, which can be used as a proxy for membrane permeability. Going forward, both soma and exchange should ideally be accounted for, if the data support the estimation of a larger number of parameters.

**Acknowledgments**

We thank reviewers, including Sune Jespersen and Jonas Olesen, for constructive comments on the manuscript. We thank Analina da Silva, Mario Lepore and Stefan Mitrea for assistance with animal monitoring. This work was made possible thanks to the CIBM Center for Biomedical Imaging, founded and supported by Lausanne University Hospital (CHUV), University of Lausanne (UNIL), Ecole Polytechnique Federale de Lausanne (EPFL), University of Geneva (UNIGE) and Geneva University Hospitals (HUG). I.O.J. is supported by the SNSF Eccellenza Fellowship PCEFP2_194260. M.P. is supported by the UKRI Future Leaders Fellowship MR/T020296/1. D.S.N. is supported by NIH under NINDS award R01 NS088040 and by the Center of Advanced Imaging Innovation and Research (CAI$^2$R, www.cai2r.net), a NIBIB Biomedical Technology Resource Center: P41 EB017183.

**Author contributions**

I.O.J., M.P. and D.S.N. designed the study; I.O.J. and F.G. conducted the study; I.O.J., A.D.S and M.P. analyzed the data; all authors wrote the paper.

**Declaration of interest:** none.

**Appendix: DKI(*t*) for the orientationally-averaged anisotropic KM**

We begin from a familiar DKI representation of a standard (scalar) KM:

$$\ln S_{KM}(b) = -bD + \frac{b^2}{2} f(1-f)(D_1 - D_2)^2 \cdot F(\tau), \quad F(\tau) = \frac{2(e^{-\tau} - 1 + \tau)}{\tau^2}, \quad \tau = \frac{t}{t_{ex}}$$

corresponding to the kurtosis (Fieremans et al., 2010) $K(\tau) = K_0 F(\tau)$ with

$$K_0 = 3f(1-f) \cdot \frac{(D_1 - D_2)^2}{D^2}.$$

We now use the above expressions for a single fascicle oriented at an angle $\theta$ to the gradient direction, as function of $x = \cos\theta$. This means that $D_1 = D_{i,\|} x^2$ and $D_2 = D_{e,\perp} + \Delta_e x^2$, where $\Delta_e = D_{e,\|} - D_{e,\perp}$. Note that $\Delta_e \equiv 0$ and $D_{e,\perp} \equiv D_e$ for the isotropic extra-cellular space assumed in the main text. To average over directions, we expand $S_{KM}(b,x)$ in moments up to $b^2$, integrate term-by-term over the orientations and re-expand in the exponential:

$$\bar{S}(b) = \int_0^1 dx\, S_{KM}(b,x) \simeq e^{-b\bar{D} + \frac{b^2 \bar{D}^2 \bar{K}}{6}}.$$

In this way, after some algebra we obtain

$$\bar{D} = \frac{1}{3}\left[f D_{i,\|} + (1-f)(3 D_{e,\perp} + \Delta_e)\right]$$

and

$$\bar{K} = K_0 F(\tau) + K_\infty,$$

where

$$K_0 = \frac{3f(1-f)\left[D_{e,\perp}^2 + \frac{2}{3} D_{e,\perp}(\Delta_e - D_{i,\|}) + \frac{1}{5}(\Delta_e - D_{i,\|})^2\right]}{\bar{D}^2},$$

$$K_\infty = \frac{4}{15} \frac{\left[f D_{i,\|} + (1-f)\Delta_e\right]^2}{\bar{D}^2}.$$

For $D_{e,\perp} = 0$ or for $f = 1$ (only sticks), $\bar{K} = K_\infty = \frac{12}{5}$, the familiar kurtosis value for the Callaghan model. Since $F(0) = 1$, the initial value

$$K(t)|_{t=0} = K_0 + K_\infty.$$

The long-time asymptotic behavior is

$$K(t)|_{t \gg t_{ex}} \simeq K_\infty + 2 K_0 \cdot \frac{t_{ex}}{t}.$$

Note that the residual kurtosis $K_\infty \equiv K(t)|_{t\to\infty}$ corresponds to that of the isotropic mixture of diffusion tensors with axial and radial diffusivities $f D_{i,\|} + (1-f) D_{e,\|}$ and $(1-f) D_{e,\perp}$, respectively.

In the main text we set $\Delta_e = 0$ and $D_{e,\perp} = D_{e,\|} = D_e$ in the above expressions, and end up neglecting $K_\infty$ when analyzing data, since the above model does not adequately describe mixing between sticks with different orientations at sufficiently long times.

**Supplementary Material** for "Neurite Exchange Imaging (NEXI): A minimal model of diffusion in gray matter with inter-compartment water exchange"

**Supplementary Methods**

DL implementation for NEXI model parameter estimation:

DL was done using an artificial neural network with three fully connected layers, with 16*(#b)*(#t) units each and leaky ReLU (rectified linear unit) activations as non-linearities. Parameters to be estimated were scaled in the [-1, 1] range and $\log(\bar{S})$ was the network input. Training was done using $10^5$ signals and their corresponding parameters, based on a loss function of mean-squared-error between the model parameter estimation and the ground truth, both for noiseless and noisy estimations. In the latter case, the SNR ranged within [80, 120] to train the network for multiple noise levels as is the case in experimental datasets.

## Supplementary Results: Figures and Tables

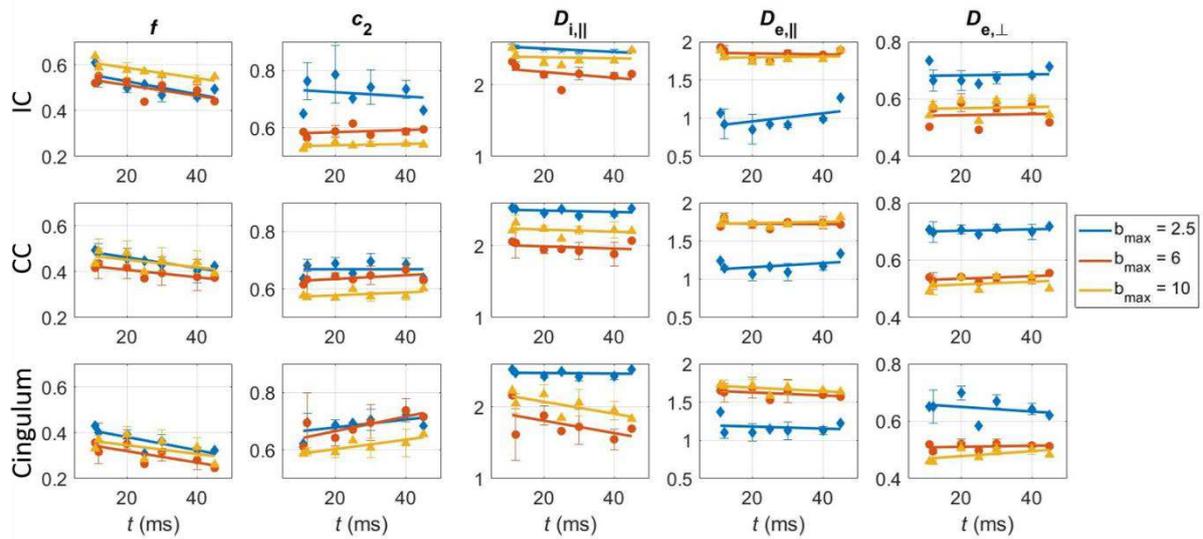

**Figure S1.** Time-dependence of Standard Model parameter estimate, also as a function of maximum b-value available, in three WM ROIs: internal capsule (IC), corpus callosum (CC) and cingulum. Symbols: mean ± std across rats. Solid line: linear fits.

| x10⁻³ | Internal capsule | Corpus callosum | Cingulum | Cortex | Hippocampus |
|---|---|---|---|---|---|
| $f$ | -2.2 ± 0.6 | -1.7 ± 1.1 | -1.9 ± 1.4 | -2.2 ± 0.9 | -2.4 ± 1.4 |
| $c_2$ | 0.24 ± 0.24 | 0.52 ± 0.39 | 1.6 ± 0.5 | 3.7 ± 0.6 | 3.1 ± 0.6 |
| $D_{i,\parallel}$ | -0.70 ± 3.1 | -1.4 ± 2.1 | -8.5 ± 3.7 | -18 ± 5 | -15 ± 8 |
| $D_{e,\parallel}$ | 0.55 ± 2.3 | 0.88 ± 1.6 | -2.4 ± 1.5 | -5.6 ± 2.4 | -9.4 ± 3.4 |
| $D_{e,\perp}$ | 0.24 ± 1.1 | 0.48 ± 0.82 | 0.91 ± 0.57 | 2.1 ± 1.1 | 3.7 ± 1.8 |

**Table S1.** Rates of change of SM parameters (mean ± σ) as a function of diffusion time (in x10⁻³ ms⁻¹ for $f$ and $c_2$, and $\times 10^{-3}\, \mu m^2 \cdot ms^{-2}$ for compartment diffusivities), for $b_{max}$ = 10 ms/μm². Light green cells: The 67% confidence interval does not cross zero: a trend in parameter change with diffusion time can be identified. Dark green cells: The 95% confidence interval (2σ) does not cross zero: a significant trend in parameter change with diffusion time can be identified. The mean slope value provides an estimate of the effect size for time-dependence, e.g. a trend of increased neurite coherence (c2) is found with longer diffusion times in corpus callosum and cingulum WM ROIs, but the rate of increase is 2 – 6x faster in GM ROIs.

| x10⁻³ | IC | CC | CG | Cortex | Hippocampus |
|---|---|---|---|---|---|
| MD | -0.05 ± 1.0 | 1.9 ± 1.1 | 0.02 ± 1.0 | -0.57 ± 0.73 | -0.06 ± 0.42 |
| MK | -2.9 ± 1.2 | -3.1 ± 0.9 | -4.8 ± 1.5 | -6.4 ± 1.0 | -7.3 ± 1.0 |

**Table S2.** Slope of linear regression (mean ± σ) for MD($t$) and MK($t$), over a $t$ = 10 – 45 ms range. Note the $10^{-3}\, \mu m^2 \cdot ms^{-2}$ units. The confidence bounds are too large around zero to conclude there is significant time-dependence of diffusivities in either ROI (note also the positive slope in CC is unphysical). The time-dependence of MK is significant for all ROIs and the slope is most pronounced for GM ROIs.

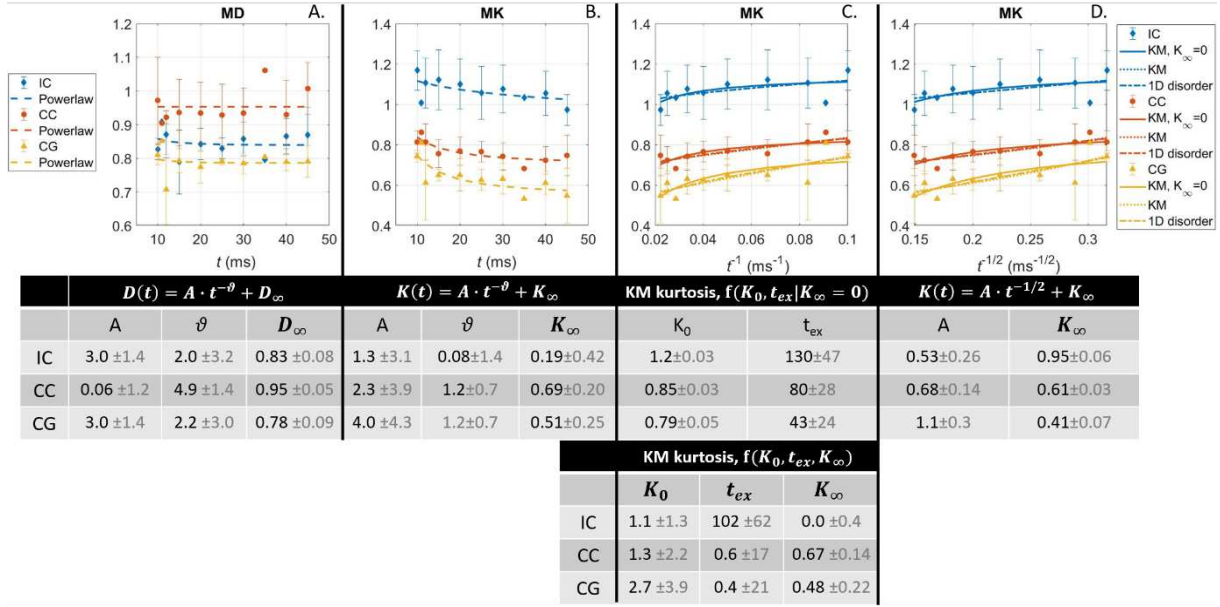

**Figure S2**. Mean diffusivity and kurtosis as a function of diffusion time, in the WM ROIs: internal capsule (IC), corpus callosum (CC) and cingulum (CG), averaged across animals. Fit parameters (mean±std) for each functional form are collected in the tables. A: Fitting the power-law to MD yielded very large coefficient α (with high variability), mainly driven by the diffusion times 10 – 20 ms. B: MK showed a decay throughout the 10 – 45 ms span. With the exception of IC, the power-law fit to MK yielded a coefficient $\vartheta$ close to 1 (and with reduced variability). C-D: The direct fitting to either the KM kurtosis (imposing $K_\infty = 0$) or the 1D structural disorder form ($\vartheta = 1/2$) showed both approaches fit the data similarly. Releasing $K_\infty = 0$ in the KM results in a similar curve to 1D disorder but with poorer parameter estimates (3 free parameters instead of 2).

| x10⁻³ | $f$ | $f_s$ | $D_{i,\parallel}$ | $D_e$ | $R_s$ |
|---|---|---|---|---|---|
| Cortex | -1.8 ± 1.4 | -3.5 ± 1.3 | -5.7 ± 3.0 | 7.9 ± 1.6 | 94 ± 12 |
| Hippocampus | -2.2 ± 1.6 | -4.8 ± 1.3 | -6.5 ± 3.4 | 8.4 ± 1.4 | 93 ± 13 |

**Table S3.** Rates of change of SANDI parameters (mean ± std) as a function of diffusion time (in x10⁻³ ms⁻¹ for $f$ and $f_s$, $x10^{-3}\ \mu m^2 \cdot ms^{-2}$ for compartment diffusivities and $x10^{-3}\ \mu m \cdot ms^{-1}$ for soma radius). Dark green cells: The 95% confidence interval (2σ) does not cross zero: a significant trend in parameter change with diffusion time can be identified, as supported by an F-test (see **Figure S9**). Light green cells: The 67% confidence interval (σ) does not cross zero.

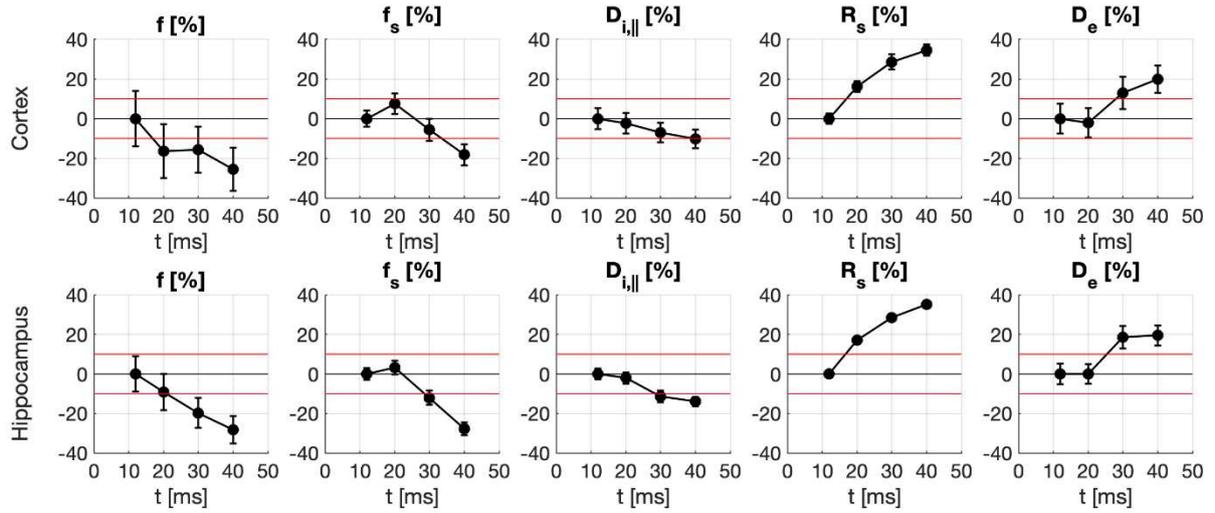

**Figure S3.** Mean percentage differences for each of the five SANDI model parameters, estimated at different diffusion times *t* with respect to the initial points at *t*=12 ms. Error bars indicate the standard error of the mean percentage difference, computed according to the error propagation formula. Red solid lines mark the ±10% region while black solid lines mark the zero mean percentage difference.

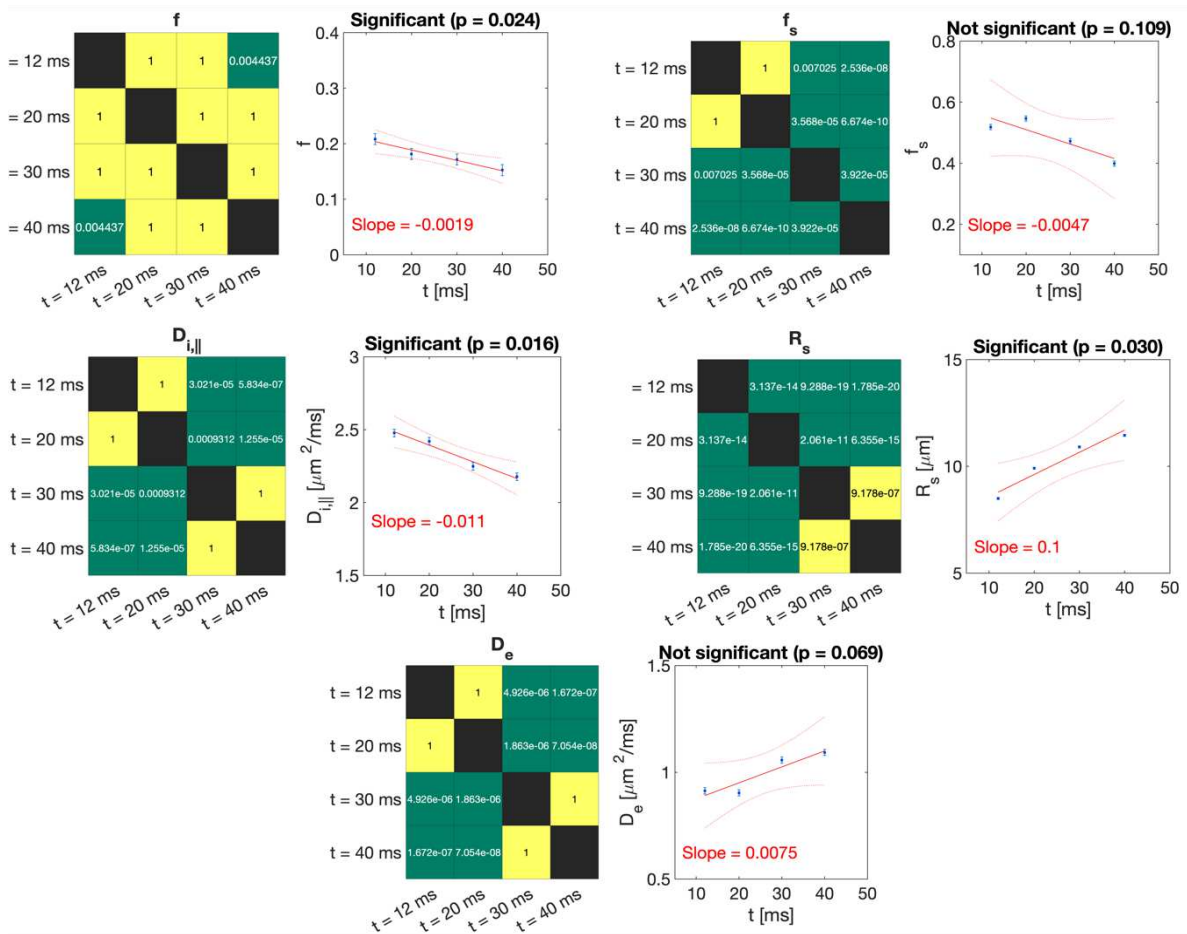

**Figure S4.** Statistical analysis of diffusion time-dependence of SANDI parameters estimates. (Left panel) Heatmaps showing which SANDI parameter estimates are significantly different between time points: green = $p<0.05$ (the exact *p*-value is reported), yellow = $p>0.05$ (a nominal *p*-value=1 is reported to help interpretability).

The results were obtained from a one-way ANOVA analysis with Bonferroni correction for multiple comparisons. (Right panel) weighted linear regression of each SANDI parameter estimate as a function of diffusion time. Solid red lines are the best linear fit, whose slope is also reported as text; dashed red lines are the 95% confidence interval. Statistically significant linear trend was assessed using an F-test on the regression model, which reports whether the model fits significantly better than a degenerate model consisting of only a constant term. The *p*-value is reported in the plot and only *p*<0.05 were considered significant.

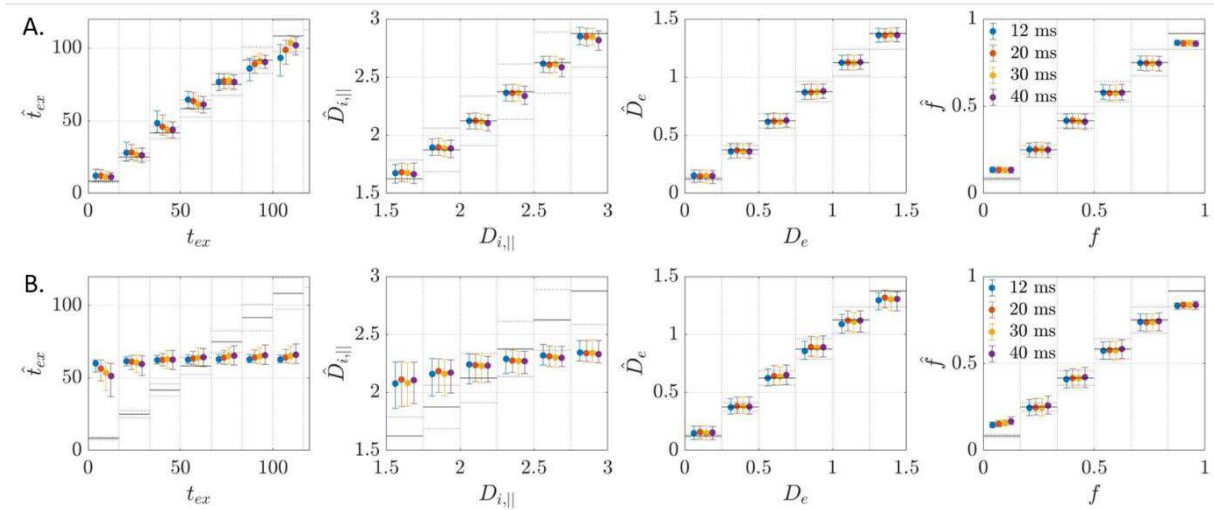

**Figure S5.** Simulation results fitting multi-shell data for each diffusion time separately using DL, without noise (A) or with SNR = 100 (B). Displayed is the ground truth (GT) vs estimation for $10^4$ set of random parameters. Markers correspond to the median & IQR in the corresponding intervals. Black lines are the ideal estimation ±10% error. In all cases, the precision is good on $D_e$ and acceptable on f. However, in a finite SNR case, $D_{i,\parallel}$ and $t_{ex}$ cannot be estimated, irrespective of the diffusion time.

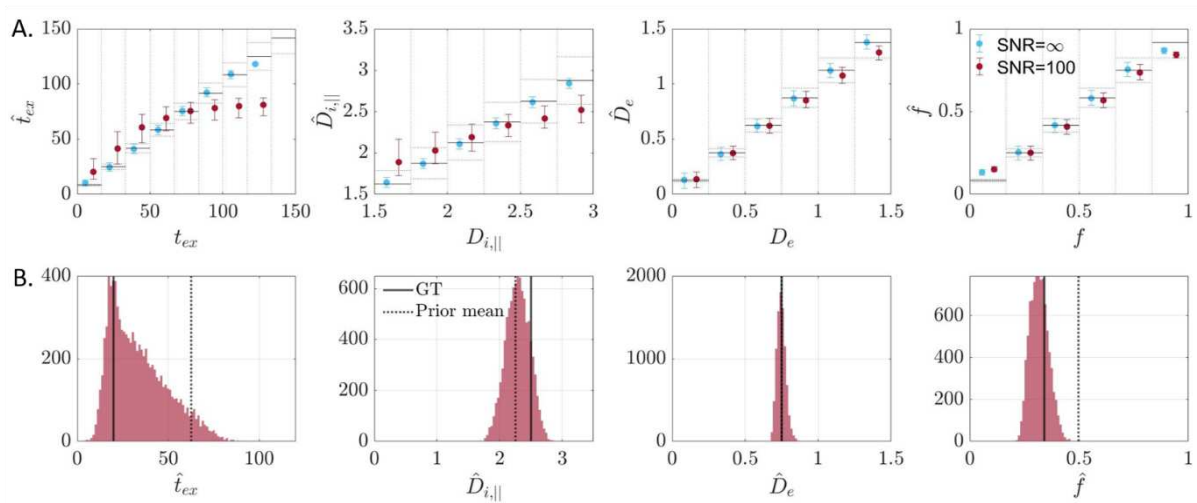

**Figure S6**. Simulation results fitting multi-shell multi-$t_d$ data jointly using DL for random GT (A) or fixed to $[t_{ex}^{th}, D_{i,\parallel}^{th}, D_e^{th}, f^{th}] = [20, 2.5, 0.75, 0.34]$ (B). **A**: Displayed are the medians & IQR in each bin. Black lines: ideal estimation ±10 % error. Without noise, DL fits all parameters with high accuracy and precision. At SNR=100, some sensitivity to $D_{i,\parallel}$ and high $t_{ex}$ values is lost but still significantly better than single $t_d$ fits (Figure S3). **B**: At SNR=100, good accuracy is achieved for $t_{ex}$, $D_e$ and f. For $D_{i,\parallel}$, DL biases the outcome towards the mean of the prior.

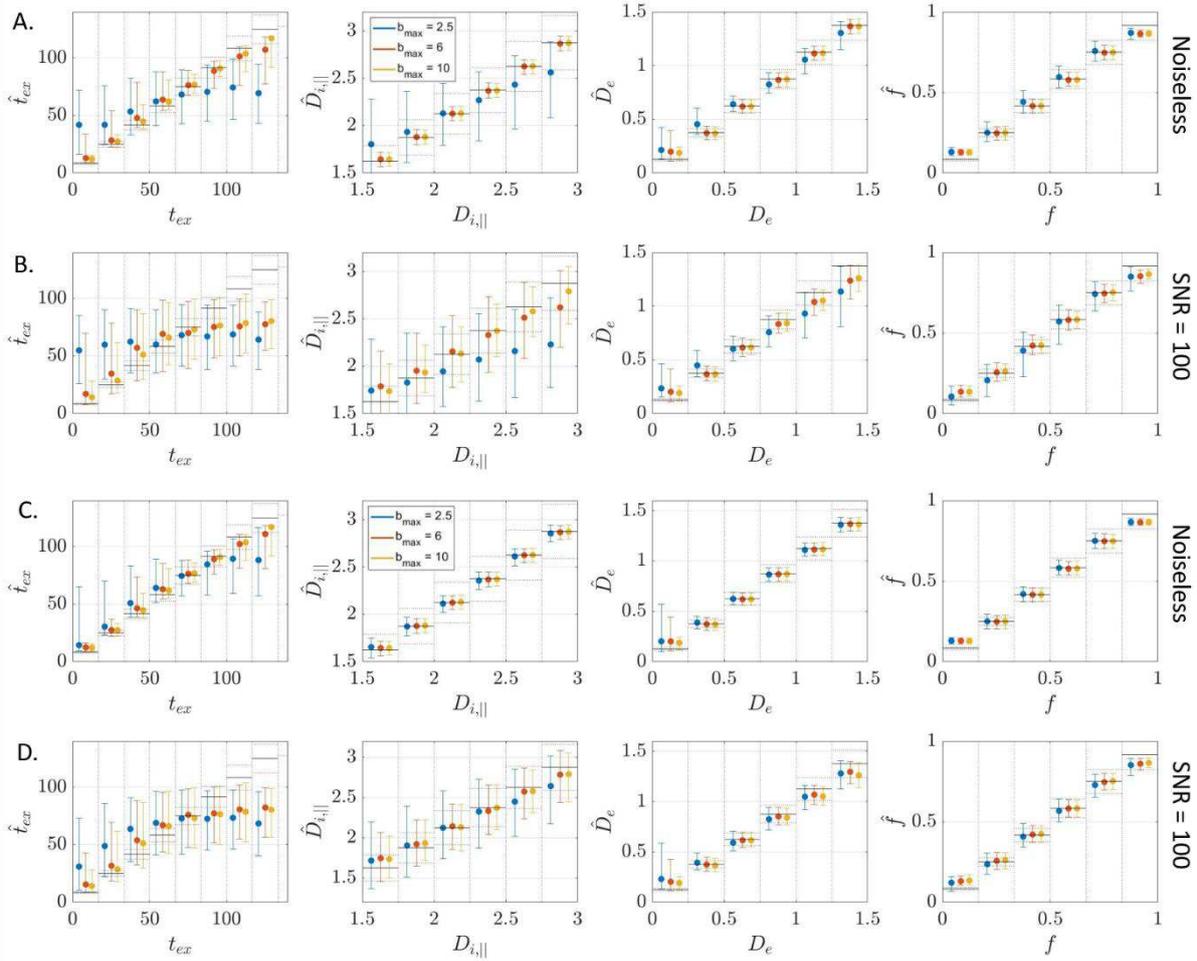

**Figure S7**. **A-B**: NEXI parameter estimates based on simulated data at four diffusion times, with increasing b-values and shells: $b_{max}$ = 2.5 (2 shells), 5.5 (4 shells), or 10 ms/µm² (7 shells). **C-D**: NEXI parameter estimates based on simulated data at four diffusion times, with increasing b-values but constant number of shells (seven): $b_{max}$ = 2.5, 5.5, or 10 ms/µm². NLLS estimation. Both the total number of shells and the maximum b-value increase the accuracy and precision of NEXI estimates. At constant number of shells, the benefit of $b_{max}$ = 10 over 5.5 is marginal.

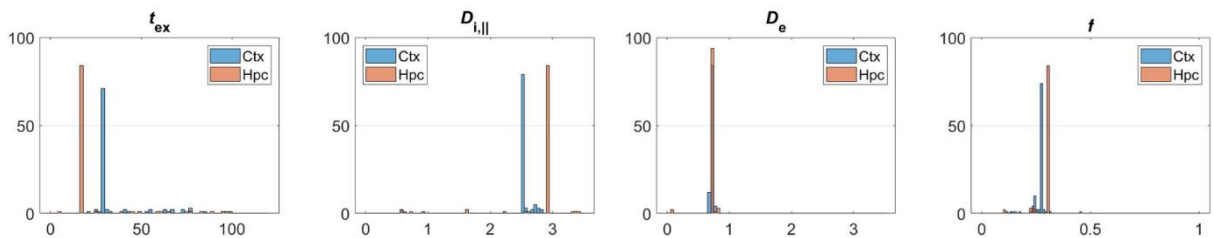

**Figure S8.** Histogram of NEXI parameter estimates in the cortex and hippocampus for N=100 random NLLS initializations. In spite of several local minima, the solution where $D_{i,\parallel} > 2$ and $D_e < 1$ is largely dominant.

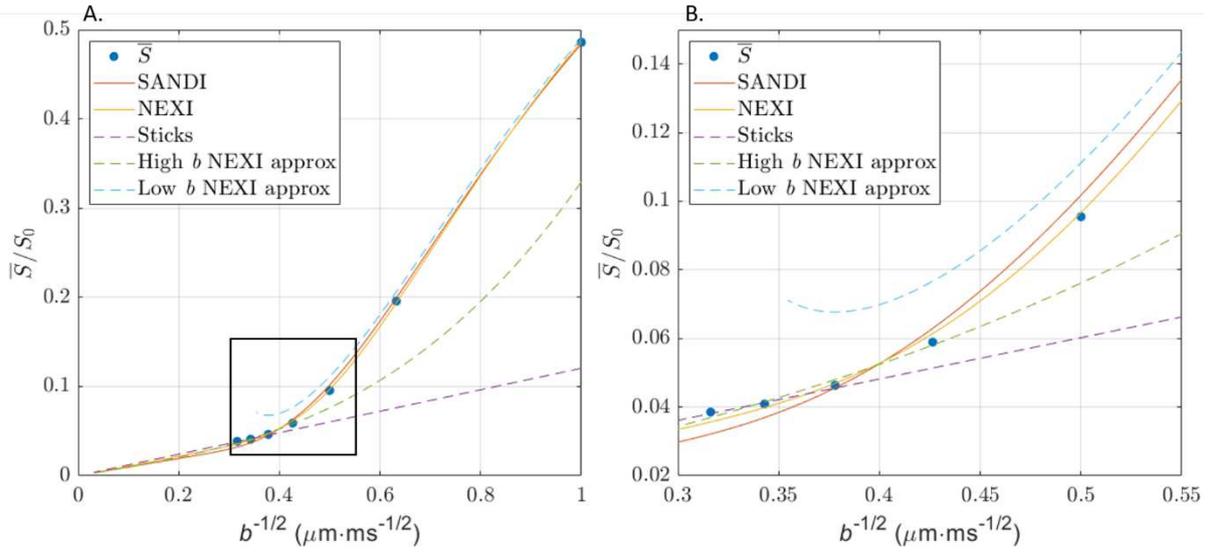

**Figure S9.** Various models were fit to the average signal in the cortex (Rat #2) at *t*=40 ms: SANDI, Eq (10), and NEXI, Eq (6), covering the full *b*-value range; the impermeable stick approximation (Callaghan's model); NEXI approximation at high *b*, Eq (8); and the NEXI-derived diffusivity + kurtosis approximation at low *b* (Appendix). **B.** Zoom-in of the black framed region in panel **A**. NEXI explains the data at t=40ms better than SANDI. Callaghan's model does not describe diffusion signal decay in the cortex appropriately due to the signal's notable curvature with respect to $b^{-1/2}$, cf Eq (8), though less pronounced that for t=12 ms. The NEXI low-b and high-b approximations are reasonable in their respective regimes. Estimated model parameters, underlying the plotted curves: SANDI: $f = 0.16$; $f_s = 0.46$; $D_{i,\parallel} = 2.3$; $R_s = 11$; $D_e = 0.46$; NEXI: $f = 0.22$; $D_{i,\parallel} = 2.4$; $D_e = 0.76$; $t_{ex} = 185$; Sticks: $f = 0.21$; $D_{i,\parallel} = 2.5$; High b NEXI approx.: $f = 0.34$; $D_{i,\parallel} = 2.2$; $D_e = 0.72$; $t_{ex} = 35$.